\documentclass[11pt,a4paper,reqno,english]{article}

\usepackage{amssymb,amsmath,amsthm}
\usepackage[pdftex]{graphicx}
\usepackage[margin=2cm]{geometry}
\usepackage[toc,page]{appendix}
\usepackage{float}
\usepackage{caption}
\usepackage{afterpage}
\usepackage{listings}
\usepackage{fancyhdr}
\usepackage{color}
\usepackage{hyperref}
\usepackage{babel,blindtext}
\usepackage[normalem]{ulem}
\usepackage{multirow}
\usepackage{lscape}
\usepackage{graphicx}
\usepackage{subfig}
\usepackage{mathtools}
\usepackage{indentfirst}
\usepackage{algorithm}
\usepackage[noend]{algpseudocode}
\usepackage{ulem}
\usepackage[toc,page]{appendix}
\usepackage{arydshln}
\usepackage{eurosym}
\RequirePackage[OT1]{fontenc}
\usepackage{authblk}

\usepackage[style=authoryear, hyperref=true, dashed=false, backend=bibtex]{biblatex}
\begin{document}

	\title{Motor Insurance Accidental Damage Claims Modeling with Factor Collapsing and Bayesian Model Averaging} 
	\author[1,2]{Sen Hu\thanks{sen.hu.1@ucdconnect.ie}}
	\author[1,2]{Adrian O'Hagan\thanks{adrian.ohagan@ucd.ie}} 
	\author[1,2]{Thomas Brendan Murphy\thanks{brendan.murphy@ucd.ie}}
	\affil[1]{School of Mathematics and Statistics, University College Dublin, Ireland}
	\affil[2]{Insight Centre for Data Analytics, Dublin, Ireland} 
	\date{}  
	\maketitle

\begin{abstract}
Accidental damage is a typical component of motor insurance claim. Modeling of this nature generally involves analysis of past claim history and different characteristics of the insured objects and the policyholders. Generalized linear models (GLMs) have become the industry's standard approach for pricing and modeling risks of this nature. However, the GLM approach utilizes a single ``best'' model on which loss predictions are based, which ignores the uncertainty among the competing models and variable selection. An additional characteristic of motor insurance data sets is the presence of many categorical variables, within which the number of levels is high. 
In particular, not all levels of such variables may be statistically significant and rather some subsets of the levels may be merged to give a smaller overall number of levels for improved model parsimony and interpretability. 
A method is proposed for assessing the optimal manner in which to collapse a factor with many levels into one with a smaller number of levels, then Bayesian model averaging (BMA) is used to blend model predictions from all reasonable models to account for factor collapsing uncertainty.
This method will be computationally intensive due to the number of factors being collapsed as well as the possibly large number of levels within factors. Hence a stochastic optimisation is proposed to quickly find the best collapsing cases across the model space.
\end{abstract}

\section{Introduction}
\label{sec:intro}

Accidental damage is a typical type risk component within motor insurance, which accounts for the damage caused to the policyholder's car by the policyholder or a named driver. In most cases it only comes as part of a comprehensive cover policy. Claim modelling of this type as well as other possible types such as third party coverage are vital in motor insurance pricing. In motor insurance, within a competitive insurance market, the main target of pricing is to charge a premium, reflecting the policyholder's risk (i.e. expected loss) accepted by the insurer, that is as accurate as possible, without customers being over-charged and potentially moving to a different insurer. More broadly, motor insurance is a typical example of general insurance (GI) where the same pricing target and the problems raised in achieving the target are equally present. 
Statistical analysis for GI pricing involves analysis of past claims history as well as different properties of the insured objects and the corresponding policyholders for the portfolio, so that relationships between key ratios (e.g. claim occurrence probability, claim frequency, claim severity, pure premium) and various rating factors (predictors) can be estimated.

Generalized linear models (GLMs) (\cite{Nelder1972glm}) have become the industry's standard approach for claim modelling. If $N$ is the number of claims, $Y$ is the expected claim size per claim and $X$ represents the characteristics of one policyholder, a standard GI pricing model would be:
\begin{equation}
\mathbb{E}(\text{Total Claim Size}|X)
= \mathbb{E}(N|X) \mathbb{E}(Y|N>0, X),
\label{eq:FrequencySeverity}
\end{equation}
where the first part of Equation~\ref{eq:FrequencySeverity}, $E(N|X)$, models claim frequency and the second part, $E(Y|N>0, X)$, models claim severity (\cite{Ohlsson2010}). 
The choice of distributions or models within the GLM framework is an extensive research field among the insurance community, see for example \textcite{Yip2005}, \textcite{Jorgensen1994}, \textcite{Smyth2002tweedie} and \textcite{Antonio2006}. In this article, a standard GLM approach is illustrated in order to focus on the proposed method, hence the Poisson and Gamma distributions are used for frequency and severity models respectively in conjunction with a log link function.  

One of the characteristics of GI claim data that adversely affects model parsimony is the presence of many categorical rating factors such as vehicle brand, home address and professional occupation. Additionally it is common that when variables are continuous, they are categorized via banding e.g. age into age intervals. This is because, firstly, there is seldom a clear linear relation between continuous variables and key ratios, so banding aids in capturing potential nonlinear effects. Secondly, it simplifies risk classification, so policyholders can be classified into risk-homogeneous groups. Thirdly, regulations often limit the way predictors can be used in ratemaking. 
Essentially, claim modelling clusters customers into risk-homogeneous groups as accurately as possible, so that within each group all customers have very similar risk profiles that can be explained via their relative ratings. 
When those factors have too many levels (either nominal or ordered), these levels should be merged to form groups due to either lack of sufficient observations for some levels or too many parameters for a standard GLM structure and for the size of the data set. Even when a model is fitted, not all categories will be statistically significant, and some should be merged to form a more parsimonious model. 
These kinds of multi-level factors represent a frequent problem within and beyond motor insurance claim modelling or actuarial science and have been widely discussed in statistical science, machine learning, and business analytics. 

Various methods have been proposed to tackle the issue: generally either to collapse all categories or to keep categories unchanged. For the former, a traditionally standard approach is pairwise comparisons within factors with Tukey's test (\cite{Tukey1949}; \cite{Hothorn2008simultaneous}; \cite{Bretz2011}) which determines which levels differ from one another simultaneously; penalisation based sparsity models (\cite{Gertheiss2010}; \cite{Tibshirani2005}; \cite{Petry2011}; \cite{She2010}; \cite{Bondell2008}; \cite{Bondell2009}) forces parameters of some categories to be equal or zero by regularisation constraints; model-based clustering (\cite{Malsiner-Walli2017}) clusters the parameters of the categories by a Bayesian MCMC framework; the ``BMA" package (\cite{BMApackage}) in R (\cite{Rsoftware}) also proposes a method where categories from all categorical variables are randomly sampled before model fitting to reduce number of model parameters. Other ad-hoc methods such as classification and regression tree (CART) (\cite{Friedman2001}) can also be interpreted as grouping categories by looking at the end nodes of the tree. 
For the latter case of keeping all categories, mixed effect models (\cite{Faraway2006extending}) can be used to treat the multi-level factors as random effects to account for their variance in the model; \textcite{Ohlsson2008GLMcredibility} proposed a method combining GLMs with credibility theory to estimate the effect of each category as a combination of the group effect and the individual level effect based on the information available for this category in the data. 
However, all methods ignore the uncertainty of the chosen clustering or chosen model and require extra paramters to be estimated in the process, except the ``BMA" package which, although it accounts for model uncertainty, fails to cluster effects of categories at the same time. 
When keeping all categories unchanged, the complexity of the model is still high.

It is almost always the case that a single final model is selected from the model space and is treated as the most representative model that may have generated the data at hand, while other potential models may have very similar properties regarding model selection criteria or predictive power. One typical scenario in claim modeling is that one variable is marginally significant, or one factor as a whole is not significant but there are a few significant categories within it. Sometimes the decision may be made based on the modeller's judgment or experience. Therefore, variable selection and categorical levels combination and selection increase the uncertainty of the selected model. 
One possible solution is, when faced by marginally significant variables, to fit two models, one of which includes the variable while the other excludes it. These models are then combined with respect to predictions.
The idea of combining models has been proposed widely in statistical literature, particularly Bayesian model averaging (BMA), which has been shown to be useful regarding model selection and improvement of model predictions (\cite{Roberts1965}; \cite{Draper1995}; \cite{Raftery1996approximate}; \cite{Hoeting1999bma} and \cite{Clyde2003model}).
Essentially, BMA and the problem of multi-level factors lead to three questions in claims modeling:
\begin{enumerate}
	\item should a categorical predictor be included?
	\item when included, should certain levels be merged together?
	\item when included and with certain levels merged, how much confidence should be placed on this clustering of levels and this model?
\end{enumerate}
Therefore, factor collapsing (FC) (or effect fusion or categories clustering) and Bayesian model averaging (BMA) are considered with the aim of improving model predictions in claim modelling, with a focus on finding the optimal manner for combining levels so that the number of levels can be reduced and thus model parsimony and interpretability improved.

The structure of this article is as follows: in Section~\ref{sec:data} data sets used are introduced;
Section~\ref{sec:factor_collapsing} briefly reviews BMA, introduces the method of FC incorporated within BMA (FC-BMA) and discusses how to select the optimal combination of categorical levels and models in a computationally efficient way;
Section~\ref{sec:small_example_benchmark} tests and compares the different approaches using a small motivating data set;
Section~\ref{sec:real_example} applies the claim models with FC-BMA in a complete example based on real industry data from a large Irish GI company.

\section{Insurance claim data}
\label{sec:data}
\subsection{Irish motor insurance data}
A large motor insurance claims data set obtained from a GI company in Ireland is examined in this article. The data represents the insurer's book of business written in Ireland over the period January 2013 to June 2014, with 452,266 policies and their characteristics included, such as policy duration, premium, claim history, policyholder demographic information, vehicle characteristics, cover options and benefits. 
In Ireland, as in most other developed countries, motor insurance is required for all motorists and is enforced by Irish law, with third party (TP) cover being the minimum requirement (\cite{MotorLaw}).
This insurer provides two different types of coverage:
\begin{enumerate}
	\item Third party, fire and theft (TPFT) cover: a combination of TP bodily injury, TP property damage and coverage for loss of the vehicle through fire or theft and any loss because of attempted fire or theft.
	\item Comprehensive cover: provides for almost every eventuality including everything in TPFT cover, as well as coverage for any damage incurred to the insured's vehicle and medical care payments regarding the registered driver's injury, regardless of how an accident occurs.
\end{enumerate} 
Across the portfolio, there are five distinct categories to be predicted separately on claims: (1) accidental damage (2) TP bodily injury (3) TP property damage (4) windscreen (5) theft. 
In this article only accidental damage (available only in comprehensive cover) is analysed, which contains 345,004 observations in total, though the same analysis could be applied across different categories.
Most of the policies (332,835) were not cancelled before the end of their policy year.  
Out of the cancelled policies, most cancelled during the first 2 months, particularly at the end of the first month.
Hence, as is common practice, period exposed to risk is allowed for in model fitting.
In the original data set, there are more than 100 variables. 27 relevant rating factors are selected in this analysis. Table~\ref{table:case_study_dataset_pickedvar} shows the predictors and their categorical levels.
Note that for data sensitivity and confidentiality reasons not all significant predictors are selected and presented.
%\begin{landscape}
	\begin{table}[h]
		\centering
		\caption{Descriptions of part of the selected Irish GI data set. ``No claim discount" represents the number of years with no claim; ``Total excess" is the amount the policyholder must pay towards any claims made on the policy including both compulsory and voluntary excesses.}
		\label{table:case_study_dataset_pickedvar}
		\resizebox{\linewidth}{!}{
		\begin{tabular}{llllllllll}
			\hline
			Variables  & Categories  \\
			\hline
%P.H. age   		  & range from 17 to 99 \\
Policyholder gender  & Female; Male; Neutral \\
Policyholder penalty point& 0; 1-2; 3-4; 5-6; 7-8; 9+  \\
%Veh. value  	  & range from 0 to 270000  \\
%Veh. age  		  & interger range from 1 to 60 \\
Vehicle fuel type 	  & Diesel; Petrol; Unknown  \\
Vehicle transmission & Automatic; Manual; Unknown \\
%Veh. engine size (cc) & range from 1000 to 5901\\
%Veh. classified group & \ \\
%Driver options    & insured only; insured \& spouse; named driver; open driver\\
Annual mileage           & 0-5000, 5001-10000, $\ldots$, 45001-50000, 50001+  \\
%Class of use      & SDP only; SDP/Retail/Publican; SDP/Trade(inc empl); SDP/Trade(inc empl,agnts); SDP/Trade (insd only) \\
Number of registered drivers & 1; 2; 3; 4; 5; 6; 7 \\
No claim discount (NCD) & 0; 0.1; 0.2; 0.3; 0.4; 0.5; 0.6\\
No claim discount protection 	  & No; Yes; Unknown \\
No claim discount stepback  	  & No; Yes \\
Total excess (\euro)   	  & 0; 125; 300; 600 \\
%M.D. licence length& \ \\
Main driver licence category& B; C; D; F; I; N \\
%Second car    	  & Higher cc; Lower cc; No second car	\\
%Product    		  & AN POST DIRECT (99,192); BROKERNET (47,358); MOTORCARE (131,384); MOTORCHOICE (67,070) \\
%%SRC channel 	  & Banca; Broker; Direct \\
%Payplan  		  & No; Yes; Yes with 0 charge  \\
%Ignition Discount & 0\%; 10\%; 10\% Curfew Only; 20\%; 20\% with curfew; 30\%; 30\% with curfew; 40\%; 40\% with curfew; IAM 20\%; P \\
%Licence FP group  & FULL-0PROV (328,326); FULL-1PROV (10,491); FULL-2+PROV (273);  PROV-0PROV (5,177); PROV-1+PROV )(737) \\
County  & 26 counties plus 4 local authrities in county Dublin  \\
\  & (Dublin City, Dun Laoghaire-Rathdown, Fingal, South Dublin); \\ 
\  & 4 cities (Limerick, Galway, Cork, Waterford), \\
\  & Unknown \\
%Mosaic Type & \ \\
			\hline
		\end{tabular} }
	\end{table}
%\end{landscape}
 
\begin{figure}[h]
	\begin{minipage}[b]{0.45\linewidth}
		\centering
		\includegraphics[width=6cm]{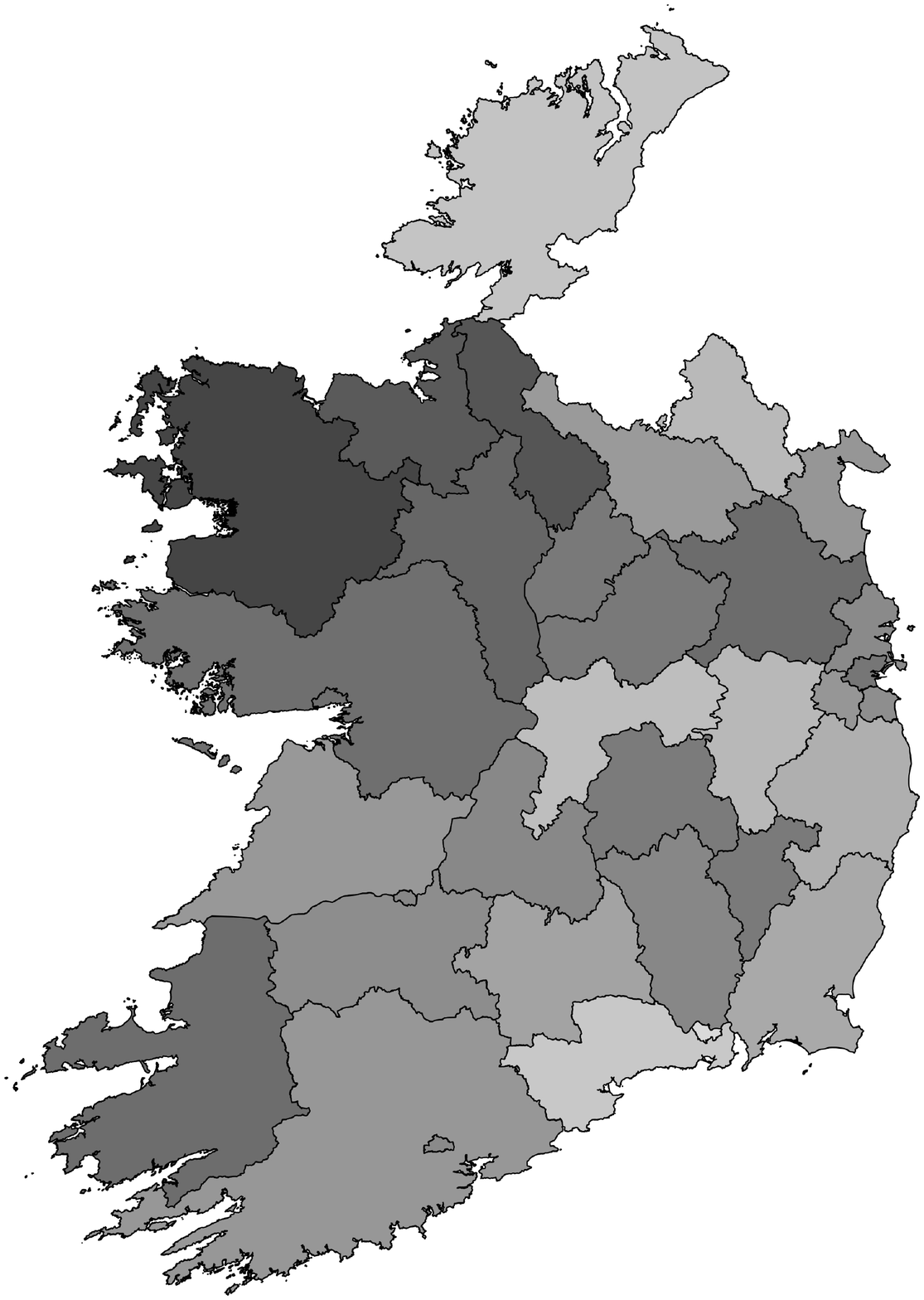}
		\caption*{(a) Frequency}
	\end{minipage}
	\hspace{0.5cm}
	\begin{minipage}[b]{0.45\linewidth}
		\centering
		\includegraphics[width=6cm]{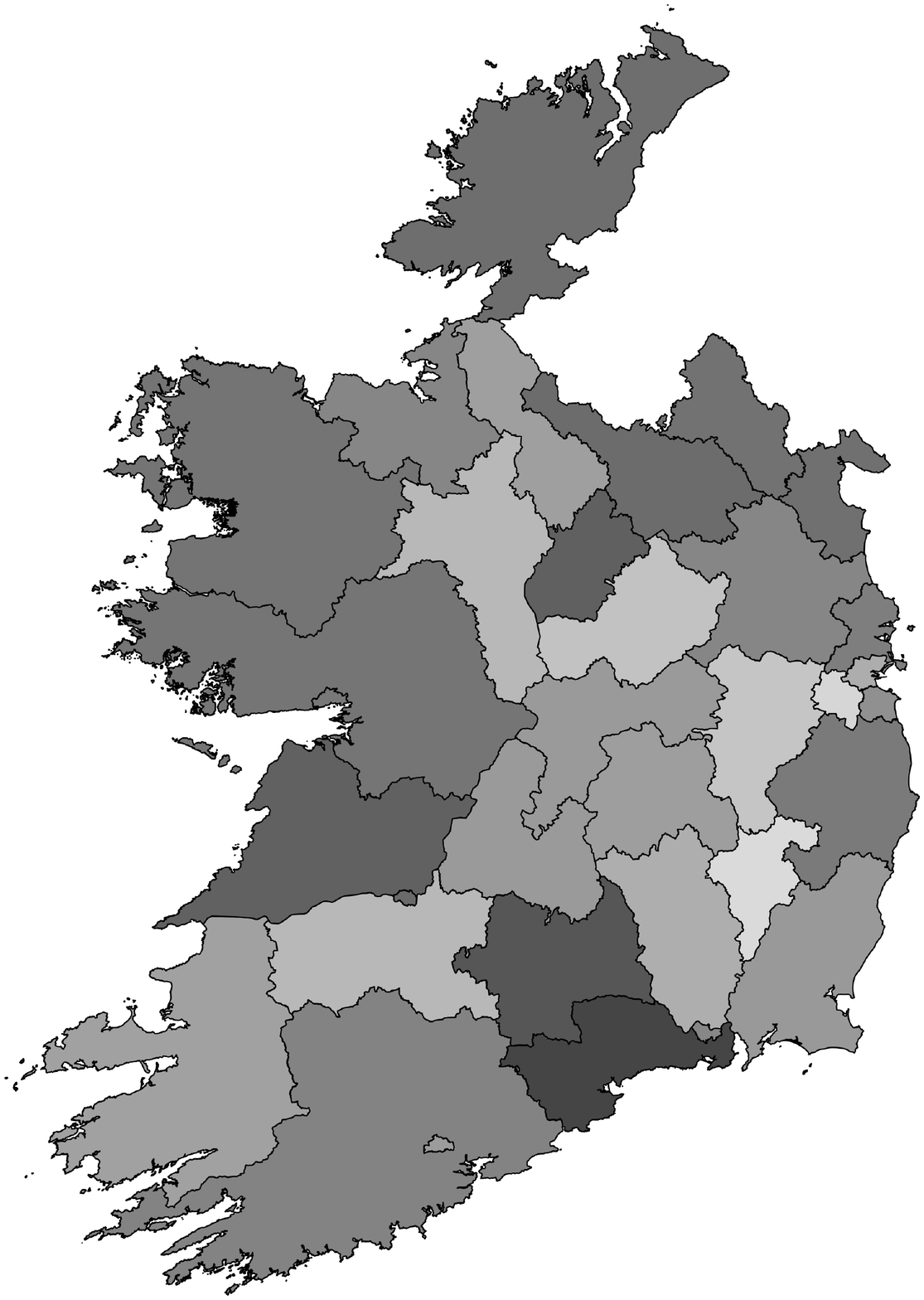}
		\caption*{(b) Severity}
	\end{minipage}
\caption{Claims distribution in Ireland by county: colors correspond to estimated coefficients when standard GLMs are fitted using all predictors for frequency and severity respectively. %from Table~\ref{table:case_study_dataset_pickedvar}. 
Fig~\ref{fig:SmallArea_Ireland_Observation_Scatter}(a) represents estimated frequency coefficients from the fitted GLM. Fig~\ref{fig:SmallArea_Ireland_Observation_Scatter}(b) represents estimated severity coefficients from the fitted GLM. The darker the color, the higher the number of claims or average claim amount in the county in question.}
\label{fig:SmallArea_Ireland_Observation_Scatter}
\end{figure}

In this article, although all factors are collapsed in the final models, we particularly focus on the county level factor showing which Irish county the policyholder lives in. There are 26 counties in total in Ireland and the Dublin region is divided into four local autorities. This variable also includes the other four main cities in Ireland (Cork, Galway, Limerick, Waterford) and an ``Unknown" category when the geo-information is missing. Hence there are 35 categories in total. 
Figure~\ref{fig:SmallArea_Ireland_Observation_Scatter} shows the claim frequency and severity levels among the counties via standard GLMs.
Further discussion on the results for this variable will be given in Section~\ref{sec:real_example}. 

\begin{figure}[h]
	\centering
	\begin{minipage}{0.48\linewidth}
		\centering
		\includegraphics[width=6cm]{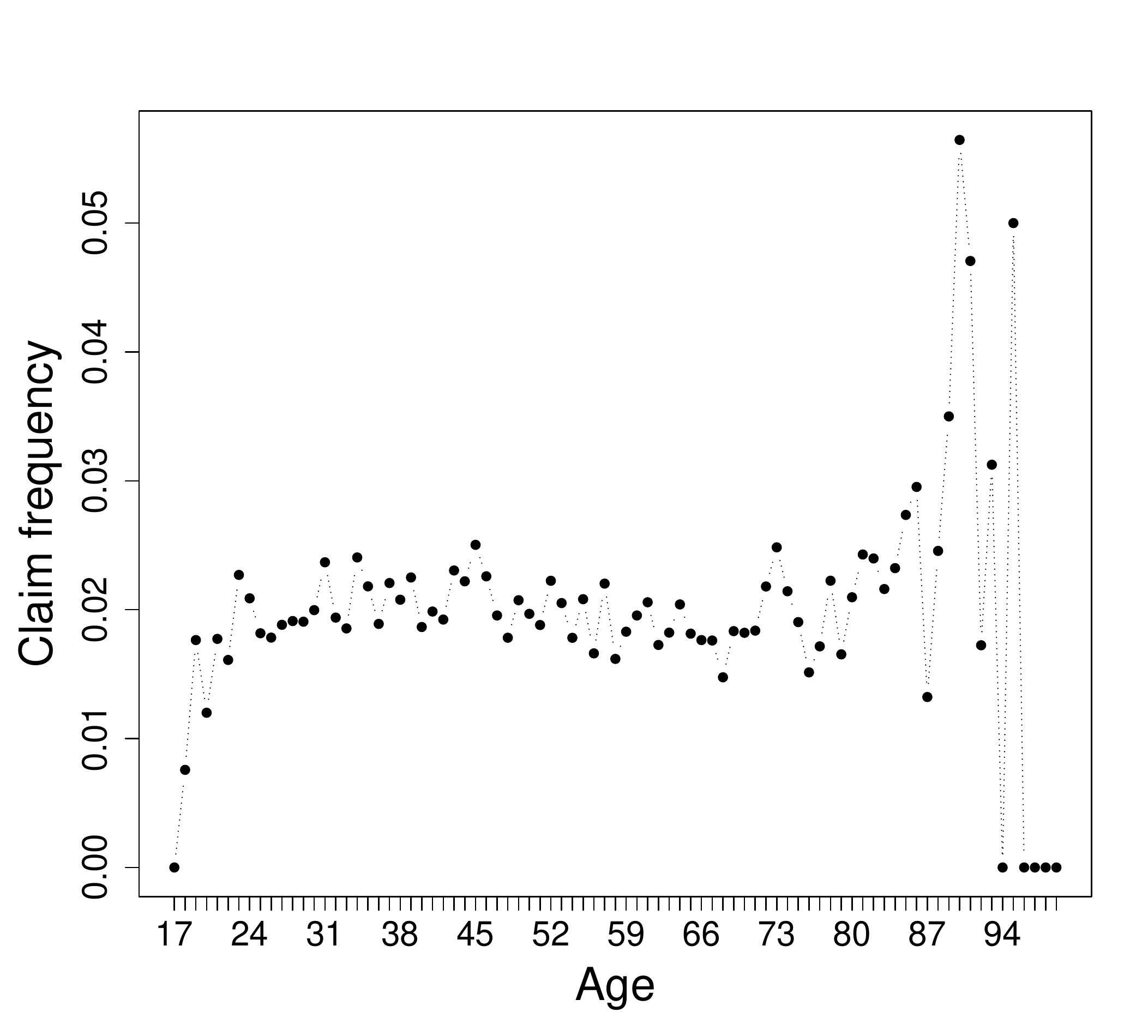}
		\caption*{(a)}
	\end{minipage}
		\begin{minipage}{0.48\linewidth}
		\centering
		\includegraphics[width=6cm]{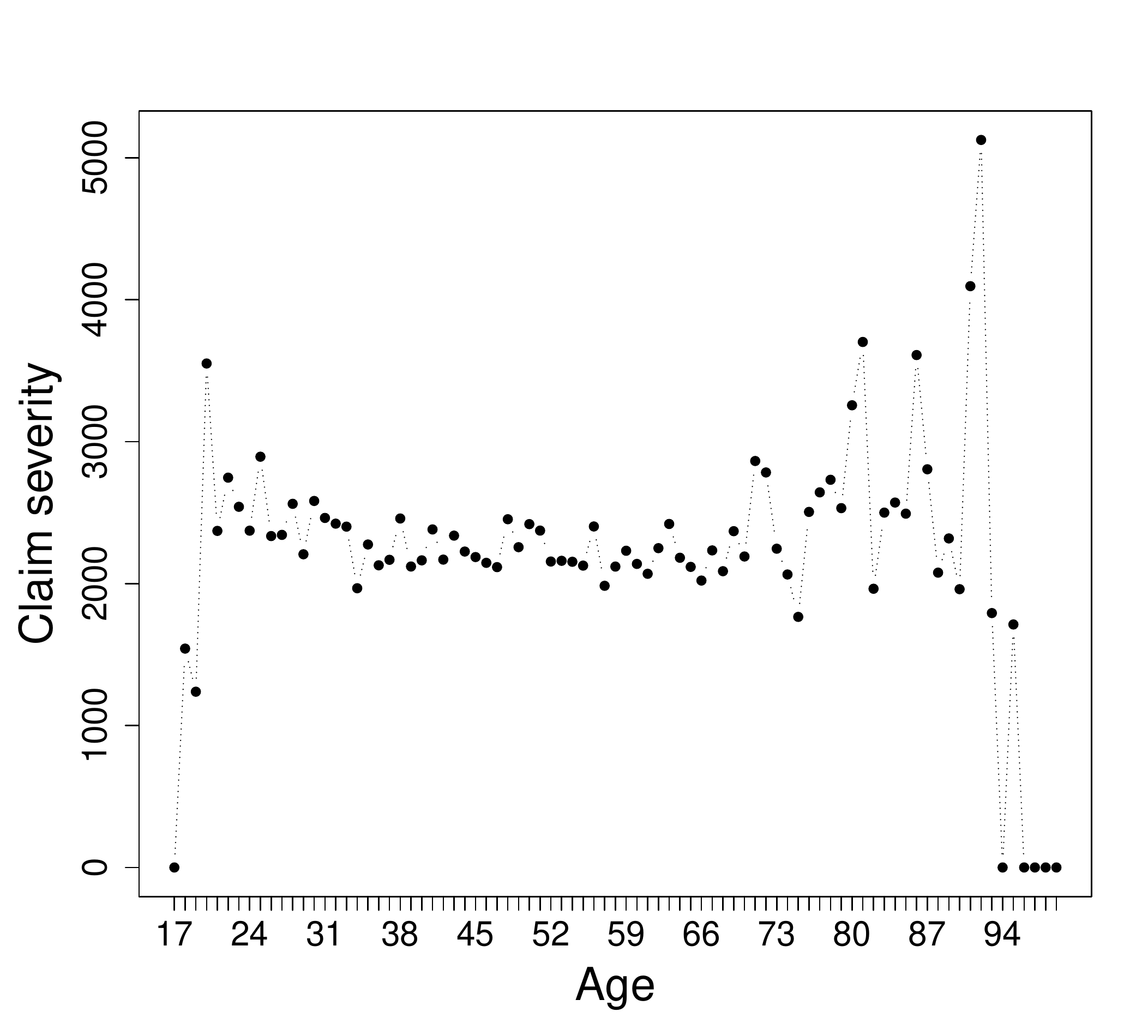}
		\caption*{(b)}
	\end{minipage}
	\caption{Claims distribution over policyholder ages: Fig~\ref{fig:claims_PHage}(a) shows average number of claims at different ages; Fig~\ref{fig:claims_PHage}(b) shows average claim sizes at different ages. }
	\label{fig:claims_PHage}
\end{figure}

There are 6,725 policies with incurred claims, accounting for about 1.95\% of the data, which means in frequency terms the vast majority of policyholders have made zero claims. 
This represents a common feature in many insurance claims data sets: semi-continuity because of the high frequency of zeroes corresponding to no claims, which has been well investigated in the insurance pricing literature (\cite{Yip2005}; \cite{Jorgensen1994}; \cite{Bermudez2012} and \cite{Shi2014}).
There is also the question of whether to treat categorical factors as nominal or ordinal: although the order of categories may contain information, keeping the order may adversely result in loss of insight for similar profiles among nonconsecutive groups. As an example, Figure~\ref{fig:claims_PHage} shows the claim distribution over different policyholder ages. It is acknowledged in insurance pricing that risk profiles of the young are similar to those of the elderly (\cite{Brown2007}; \cite{Clijsters2015}). This is reflected in Figure~\ref{fig:claims_PHage}: in Figure~\ref{fig:claims_PHage}(a) the young ($\leq 25$) and the elderly ($\geq 50$) make a similar number of claims, although when the age is over 70, the frequency becomes first much higher and then much lower, which is primarily because of fewer observations for these age groups. A similar trend is shown in Figure~\ref{fig:claims_PHage}(b) where both the young ($\leq 25$) and the elderly ($\geq 65$) have higher average claim sizes. This is complemented by the fact that in insurance pricing the relationship is generally not linear and hence categorical predictors are preferred.  

\subsection{Sweden third party motor insurance claims data}
A well-known data set called ``motorins" from \textcite{Faraway2006extending} is used to illustrate the method before applying the proposed FC-BMA method to the Irish insurance data set in Section~\ref{sec:real_example}. It contains claims history of third party (TP) motor insurance in Sweden in 1977 for 1797 observations, with combinations of 4 rating factors: Kilometers (kilometers per year, 5 levels), Make (different car models, 9 levels), Zone (geographical areas in Sweden, 7 levels) and Bonus (no claims bonus i.e. number of years since last claim filed, 7 levels).  
At that time in Sweden all motor insurance companies applied identical categorisation variables to classify customers, thus their portfolios and their claims statistics could be combined. Hence the data is in a grouped, aggregated format.
That is, each row represents one distinct type of policyholder and it also includes aggregated earned exposures i.e. total insured years (Insured), number of claims (Claims) and total losses (Payments) of policyholders.
A detailed description of the data set can be found in \textcite{Faraway2006extending} and \textcite{Smyth2002tweedie}.
Note that the factors Kilometers and Bonus are both ordered; treatment of these variables as nominal or ordinal will be discussed later.

\section{Factor collapsing with Bayesian model averaging}
\label{sec:factor_collapsing}
\subsection{Bayesian model averaging (BMA)}
\label{sec:BMA}
The problem that the standard GLM claim modelling approach faces in terms of identifying a single ``best" model is that it ignores model uncertainty i.e. how confident we should be in the selected model. 
Consequently, uncertainty about quantities of interest may be underestimated. An alternative method is to select a group of ``best" models from the model space and combine them (via either prediction or model coefficients) based on their model probabilities. One example of this application is when there is a rating factor that is marginally significant in the model, as previously mentioned.
Combining models has been investigated widely in the literature and Bayesian model averaging is a common method of combining models that involves selecting models using posterior model probability. See for example \textcite{Roberts1965}, \textcite{Draper1995}, \textcite{Hoeting1999bma} and \textcite{Raftery1996approximate}.  
Suppose $\Delta$ is the quantity of interest, such as a prediction on a new policy for claim size or rating relativities. Its posterior distribution given observations $D$ is 
\begin{equation}
Pr(\Delta|D) = \sum_{k=1}^{K} Pr(\Delta|M_k, D) Pr(M_k|D),
\end{equation}
where $M_1, \ldots, M_K$ are all selected models in the model space.
This is an average of the posterior distribution $Pr(\Delta|M_k, D)$ under each of the models $M_k$ considered, weighted by their posterior model probability $Pr(M_k|D)$.  
For the GLMs, marginal likelihood cannot be obtained by analytic integration but can be approximated by Bayesian Information Criterion (BIC) through Bayes factors (\cite{Raftery1996approximate}; \cite{Kass1995}). 
Therefore, it enables an easy approximation of posterior model probability $P(M_k|D)$, through which BMA is implemented:
\begin{equation}
P(M_k|D) \approx \frac{\exp(-0.5 BIC_k) Pr(M_k)}{\sum_{r=1}^{K} \exp(-0.5 BIC_r) Pr(M_r) } .
\end{equation}
This leads to a key decision as to the prior used. 
A flat prior is often used corresponding to prior information that is objective among competing models. 
It has been shown in the literature that noninformative priors yield satisfactory performance and are easy to explain but their use has also been criticized
(\cite{Clyde2003model}; \cite{Clyde2004} and \cite{Hoeting1999bma} and its discussion). Other informative priors could also be considered, particularly priors incorporating other information such as number of observations for each category or how many levels should be retained. One of the potential choices would be a dilution prior which considers model or prediction correlation while taking the number of observations of each category in a collapsed factor into account, see for example \textcite{George2010} and \textcite{Garthwaite2010}. 
When noninformative priors are used, the posterior model probability $P(M_k|D)$ is 
\begin{equation}
P(M_k|D) \approx \frac{\exp(-0.5 BIC_k)}{\sum_{r=1}^{K} \exp(-0.5 BIC_r) } .
\end{equation}
It is worth noting that BIC has the consistency property that it is guaranteed to select the true model as the number of observations becomes infinitely large. Hence, it has a flexible significance threshold that makes significant parameter inclusion more stringent, unlike AIC or p-values. A common feature in many insurance claims data sets is the large sample size. Therefore, it justifies the use of BIC with respect to model selection criteria. 

Another issue with BMA is how many models should be selected in the model space to be averaged over, especially when the size of model space is very large. 
One widely used method is Occam's window (\cite{Madigan1994model}), which greatly reduces the number of models in the summation. 
One method suggested in \textcite{Volinsky1997} and \textcite{Madigan1995} uses the Markov Chain Monte Carlo model composition (MC$^3$) to rapidly identify models. It constructs a Markov chain within the model space 
%(when properly defines a neighbourhood of a model, a transition matrix etc.)
, then simulates it to draw observations (i.e. models) $M_1, M_2, \ldots, M_N$. This method is a special case of Metropolis-Hastings algorithm (\cite{Chib1995}). 
%It is worth noting that the proposed stochastic process searching for the best collapsing of factors in this article is similar to this method.

It has been shown that BMA provides improved predictive performance versus using a single best model (\cite{Madigan1994model}), although the magnitude of the improvement varies.
Even though BMA has been widely used across disciplines, the authors note that BMA has not been applied in the field of actuarial claim modelling.
Predictions from all selected models are combined by taking a weighted average across model predictions. The weights are model posterior probabilities that are approximated using BIC. For simplicity, only a flat prior is considered. 

\subsection{Factor collapsing (FC)}
The mathematical concept of a set partition is used as the basis of this method: a partition of a set $\mathcal{X}$ is a disjoint collection of non-empty subsets of $\mathcal{X}$ whose union is $\mathcal{X}$ (\cite{Halmos1974}). 
If there is a collection of finite nonempty subsets $A_1, A_2, \ldots$ of $\mathcal{X}$, the sets $A_i$ are pairwise disjoint (i.e. $A_i \bigcap A_j = \emptyset$ for $i \neq j$) and the union of all $A_i$ is $\mathcal{X}$ (i.e. $A_1 \bigcup A_2 \bigcup \ldots = \mathcal{S}$), then the collection of $A_1, A_2, \ldots$ is a partition of $\mathcal{X}$. 
It is obvious that the more elements there are in the set, the more partitions there are in the set. 
The Bell number $B_n$ is defined to record the number of ways to partition a set of $n$ elements (\cite{Bell1934}). 
The Bell number can be calculated using an exponential generating function $B(n) = \sum_{m=0}^{\infty} \frac{B_m}{m!} x^m  = e^{e^n - 1}$, and the series of asymptotic expansions of $\frac{\log B_n}{n}$ has been proven as convergent for sufficiently large values of $n$ (\cite{DeBruijn1970}), which shows that the number increases super-exponentially as more elements are included in the set. As the number of elements increases, the number of possible partitions will eventually be too large and too computationally intensive to calculate every posterior model weight, a phenomenon discussed in later sections.
It is also worth noting that since unordered partitions are used, the order of partitions or elements within each partition does not matter.

For example, to collapse the rating factor Kilometers from the Sweden Third Party data set, this factor has 5 levels of kilomotres per year: $<$1k, 1k-15k, 15k-20k, 20k-25k and $>$25k, which are represented by the numbers $1, 2, \ldots, 5$ respectively.
There are 52 ways of collapsing this factor into one with a lower number of levels as in Table~\ref{table:example_factor_collapsing}. 
One example is $\{ \{1\}, \{23\}, \{45\} \}$. 
From now on, for simplicity, the notation $(1)(23)(45)$ is used to represent a set of sets instead of the proper mathematical representation. 
Hence for $(1)(23)(45)$ the new levels are $<$1k, 1k-20k and $>$20k.
A combinatorics term called graycode (or restricted growth string) is also used to record the combination (\cite{Stanton1986}; \cite{Ruskey1994}).
%It means that each partition has its blocks listed in increasing order of smallest element, such that block 1 contains element 1, block 2 contains the smallest element not in block 1, and so on. 
It is a string $a[1...n]$ where $a[i]$ is the block in which element $i$ occurs.
The graycode of $(1)(23)(45)$ is $12233$. 

\begin{table}[h]
	\centering
	\caption{The rating factor Kilometers has five levels, therefore $B_5=52$ different combinations, some of which are shown in this table. For each combination, different formats are shown: graycode, set of sets and grouping descriptions. Posterior model probabilities are also shown for each case based on the BIC values. The sum of all posterior probabilities is one. Note that the model space is not large, hence all models are selected for BMA. This example is based on the severity model.}
	\label{table:example_factor_collapsing}
	\resizebox{\linewidth}{!}{
	\begin{tabular}{c|cllccccccccc}
		\hline \hline
		\multicolumn{5}{l}{Rating factor: Kilometers} \\
		\multicolumn{5}{l}{Levels: $<$1k; 1k-15k; 15k-20k; 20k-25k; $>$25k} \\
		%\multicolumn{5}{l}{\  \ } \\
		\hline
		Index & Graycode & Grouping & Description & BIC &Model weight\\
		\hline
1 &11111& (12345)   & any number of kilometres &1878661  &0 \\     
2 &11211& (1245)(3) & $<$15k and $>$20k; 15k-20k &1878667  &0  \\    
3 &11121& (1235)(4) & $<$20k and $>$25k; 20k-25k &1878539 &0    \\  
\vdots&\vdots    & \vdots & \vdots & \vdots \\
40 &12233& (1)(23)(45)& $<$1k; 1k-20k; $>$20k  &1878161 &0.8124\\
41 &12323& (1)(24)(35)& $<$1k; 1k-15k and 20k-25k; 15k-20k and $>$25k  &1878266 &0 \\
42 &12341& (15)(2)(3)(4)& $<$1k and $>$25k; 1k-15k; 15k-20k; 20k-25k;  &1878402 &0      \\
\vdots&\vdots    & \vdots & \vdots & \vdots \\
46 &12342& (1)(25)(3)(4) & $<$1k; 1k-15k and $>$25k; 15k-20k; 20k-25k;  &1878198  & 0      \\
47 &12234& (1)(23)(4)(5) & $<$1k; 1k-20k; 20k-25k; $>$25k &1878167 & 0.0379 \\
48 &12324& (1)(24)(3)(5) & $<$1k; 1k-15k and 20k-25k; 15k-20k; $>$25k &1878225& 0      \\
\vdots&\vdots    & \vdots & \vdots & \vdots \\
50 &12334& (1)(2)(34)(5)  & $<$1k; 1k-15k; 15k-25k; $>$25k & 1878241 &0      \\
51 &12344& (1)(2)(3)(45)  &$<$1k; 1k-15k; 15k-20k; $>$20k & 1878164 &0.1430\\
52 &12345& (1)(2)(3)(4)(5)&$<$1k; 1k-15k; 15k-20k; 20k-25k; $>$25000 & 1878170 &0.0067\\
\hline\hline
	\end{tabular} }
\end{table}

When there are multiple rating factors in the model, as is normally the case, instead of collapsing each factor individually, collapsing multiple rating factors simultaneously may work better. Each combination of partitions of all factors is checked from the model space and the best combination is selected. 
This will lead to a more accurate result since different ways of collapsing factors change the covariance structure among rating factors and hence the model fitting is changed. The best partition for one factor will probably be different from the best partition for this factor when collapsed simutaneously with other factors.

\subsection{Factor collapsing with Bayesian model averaging (FC-BMA)}

For a factor with a certain number of levels, different combinations of levels using set partition are checked by fitting it in the pre-specified model (with all other aspects of the model such as other rating factors, distributions and link function unchanged) and recording some model selection criteria, such as likelihood, BIC or AIC.  
Based on the selection criteria, some of the optimal combinations will be chosen. In this way, a rating factor with many levels will be collapsed into one with a smaller number of levels, where each grouping of factors represents greater homogeneity of risk and is statistically significant. BMA can then assess the posterior probability for each selected combination and average the model predictions across all selected models. Note that only model predictions are combined to assess the FC-BMA efficiency. 
Although theoretically model parameters can also be averaged, caution is needed regarding reference levels in GLMs and the averaged coefficients will become less clustered, which essentially goes against the idea of factor collapsing. 
Table~\ref{table:example_factor_collapsing} also includes model weight, which represents posterior model probabilities, approximated by BIC.

\begin{figure}[ht]
	\centering
	\begin{minipage}[h]{0.3\linewidth}
		\centering
		\includegraphics[width=3.5cm]{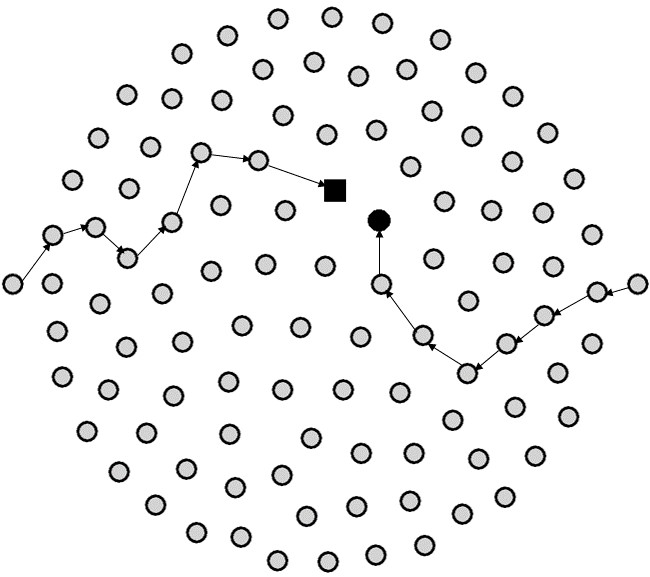}
		\caption*{(a)}
	\end{minipage}
	\begin{minipage}[h]{0.3\linewidth}
		\centering
		\includegraphics[width=3.5cm]{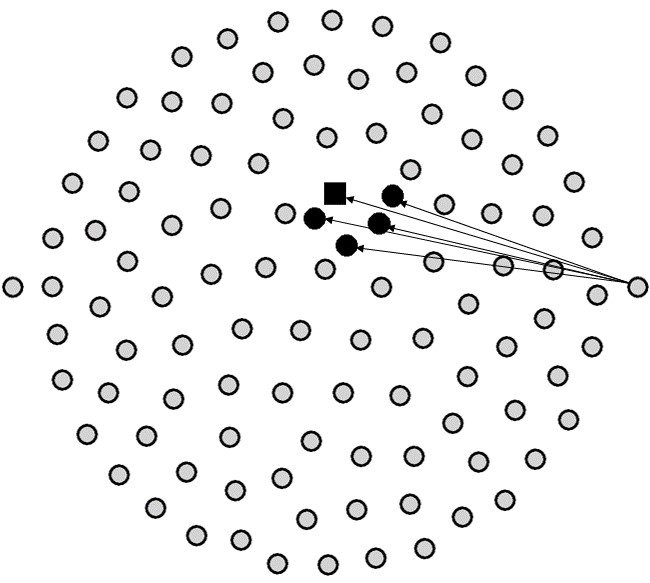}
		\caption*{(b)}
	\end{minipage}
	\begin{minipage}[h]{0.3\linewidth}
		\centering
		\includegraphics[width=3.5cm]{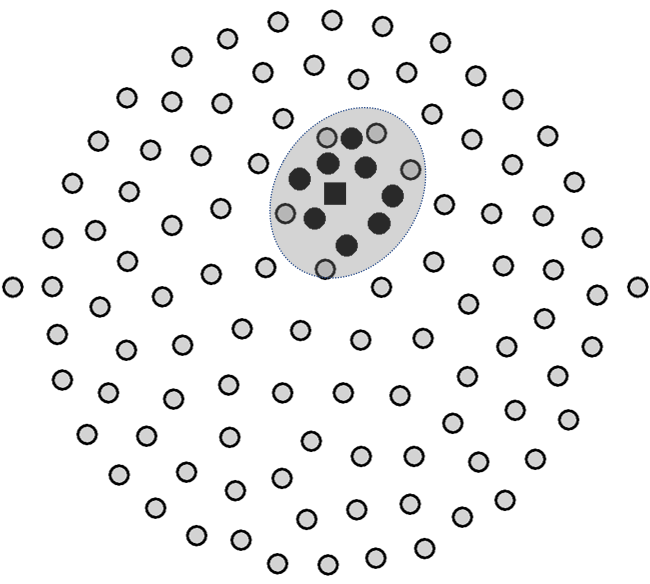}
		\caption*{(c)}
	\end{minipage}
\caption{Comparing stepwise selection methods with FC-BMA: each dot represents a model within the model space and the black square represents the global optimum model (either with or without factors collapsed). The most leftward and rightward dots are the null model and the saturated model respectively. In Fig~\ref{fig:comparison_stepwise_selection}(a) both forward and backward selection methods do not lead to the same final model, but the two models can be considered as some variation of the global optimum model (if not already equal to the global optimum model).
In Fig~\ref{fig:comparison_stepwise_selection}(b) FC-BMA not only considers variable inclusion or exclusion using the same criteria but also expands the model space by taking categorical levels selection into consideration. It finds many closely optimal models in one step.
In Fig~\ref{fig:comparison_stepwise_selection}(c) by using FC-BMA, many models from the optimal model region can be found and be averaged. They all have relatively similar fit and are very close to the global optimal model. }
\label{fig:comparison_stepwise_selection}
\end{figure}

When combining all levels within one factor, all observations in the data will have the same information for this factor, hence it is excluded from the model. This addresses the problem of variable selection.
In particular, when the selection criterion is BIC, this method is similar to standard stepwise model selection using BIC, as shown in Figure~\ref{fig:comparison_stepwise_selection}.
Starting from a null model and saturated model respectively, forward selection and backward selection adds or eliminates one whole categorical variable consecutively from the previous steps (not distinguising between its categories). As a general phenomenon, both selection methods do not always lead to the same conclusion. It can be easily envisaged that, within the model space, there is an optimal model region and conclusions from both selection methods may fall within this region where both models can be considered optimal and are both similar to the true global optimum model in the model space. By considering not only variable selection but also factor level selection (FC) using the same criteria, the model space is massively expanded. FC-BMA can then find many models within the optimal model region in one step that are similar variations of the global optimum model. 

\subsection{Conditions on set partition}
In the example above in Table~\ref{table:example_factor_collapsing}, the variable Kilometers is treated as nominal.
It can also be treated as ordinal, maintaining the internal order of levels. 
This is a typical issue in categorical analysis: how to treat the categorical levels to best suit the question at hand (\cite{Agresti2002}).
For the Irish insurer data set in Figure~\ref{fig:claims_PHage}, policyholder's age is regarded as nominal. Throughout this article categorical variables will be treated as nominal.
However, this neglects the inherent order and hence might potentially lead to some degree of loss of information. Alternatively, the order can be kept so that only adjacent levels are combined together by setting consecutive conditions on set partition. The number of ways of forming partitions in this case will be the same as the number of compositions of the integers of all elements, and will be greatly reduced.
Either case may lead to confusing results that cannot be interpreted directly. Therefore, caution should be take when deciding which condition to use. 

Other conditions can be pre-specified for set partition calculations such as that certain levels must be together, or must not be together. This is particularly important in actuarial science when an experienced pricing actuary may have insight on potential risks from past experience. In this way, the model space can be greatly reduced. 
For example, when there are 8 levels and we restrict levels 3 and 4 not to be together, then the number of models in model space reduces from 4140 to 3263. If levels 3 and 4 must be together all the time, then the number of models reduces from 4140 to 877. If we mandate that levels 2 and 3 are together, levels 4 and 7 are apart, then there are only 674 models left to be explored.

\subsection{Stochastic Optimisation}
\label{sec:stochastic_search}

\begin{figure}[H]
	\centering
	\includegraphics[width=9cm, height=6cm]{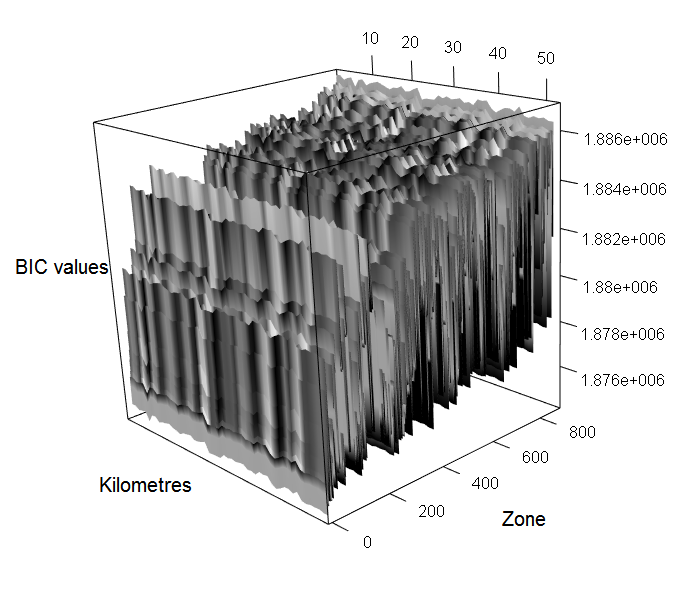}
	\caption{A 3D surface illustration of the objective function when collapsing two factors (Kilometers and Zone) over a model space of size $45,604$, from Sweden Third Party motor insurance data example. The objective function is a discontinuous BIC surface. The aim is to find a value that minimizes it. Kilometers and Zone have five and seven levels respectively, hence $52*877=45,604$ ways of being partitioned.}
	\label{figure:3dBICplot}
\end{figure}

An immediate issue for implementation of factor collapsing is that, because of the super-exponential increase in the Bell number, FC with a complete exhaustive search becomes increasingly computationally intensive, hence it is only suitable for a reasonable number of levels, perhaps less than 15. When the number of categories within a factor is over 20, the number of possible models is greater than $10^{15}$ for that factor. It is often the case in GI claim data that some rating factors contain more than 20 levels. Hence a stochastic optimisation is considered.
Given a model space that consists of all combinations of partitions over all factors, this can be regarded as an optimisation problem, in which the objective function is a discontinuous, non-differentiable and highly nonlinear surface as shown in Figure~\ref{figure:3dBICplot} and the aim is to find the global minimum of the selection criteria (e.g. BIC) across the model space.

Many stochastic optimisation techniques have been developed for dealing with highly nonlinear problems similar to the one at hand. Two methods have been implemented to address this optimisation problem: simulated annealing (SA) and the genetic algorithm (GA).

\subsubsection{Simulated annealing}
This is a common meta-heuristic method for optimisation using a Metropolis Monte Carlo simulation, originally proposed by \textcite{kirkpatrick1983optimization}. It is an adaptation of the Metropolis-Hastings algorithm, hence it shares similarities with the the Markov Chain Monte Carlo model composition (MC$^3$) approach in
\textcite{Volinsky1997}, \textcite{Madigan1995} and \textcite{Hoeting1999bma} to search over the model space to find best models. It has been shown to greatly improve computation speed while providing accurate results. 
A key component in SA is how to define transition jumps in model space. A random step is defined by randomly picking an element in the set, to put either in a different group or put aside in a new group. An example of this method is presented in Table~\ref{table:partition_neighbour_example}. Alternatively, other transitions can be defined, such as random spliting of one cluster or random grouping of two clusters in each step. 
%When collapsing multiple factors, at each iteration one variable is picked and changed to one of its neighbours. However, the probability of being picked is based on the number of levels, so a variable that has more levels would be chosen more often to explore its model space.
To implement SA, an annealing schedule needs to be selected that defines a decreasing set of temperatures, as well as starting temperature and the number of iterations at each temperature. 
There has been extensive discussion on setting simulated annealing parameters, see for example \textcite{Nourani1998}, \textcite{Lundy1986}, \textcite{Dowsland1993}, \textcite{Rayward-Smith1996} and \textcite{VanLaarhoven1987}.

\begin{table}[H]
	\centering
	\caption{Example of neighbouring partitions of a synthetic given partition, obtained by randomly moving one element from the original state.}
	\label{table:partition_neighbour_example}
	\begin{tabular}{cc|cccc}
		\hline
		\multicolumn{2}{c}{Original state} & \multicolumn{2}{|c}{Neibouring states} \\
		\hline
		Graycode & Partition    & Graycode & Partition \\
		12345 & (1)(2)(3)(4)(5) & 11234 & (12)(3)(4)(5)\\
		\     & \               & 12134 & (13)(2)(4)(5)  \\
		\     & \               & 12314 & (14)(2)(3)(5)   \\
		\     & \               & 12341 & (15)(2)(3)(4) \\
		\     & \               & 12234 & (1)(23)(4)(5)\\
		\     & \               & 12324 & (1)(24)(3)(5)\\
		\     & \               & 12342 & (1)(25)(3)(4)\\
		\     & \               & 12334 & (1)(2)(34)(5)\\
		\     & \               & 12343 & (1)(2)(35)(4)\\
		\     & \               & 12344 & (1)(2)(3)(45)\\
		\hline
	\end{tabular}
\end{table}

\subsubsection{Genetic algorithm}

The genetic algorithm (GA) is another commonly used meta-heuristic algorithm, which mimics processes observed in natural selection that drive biological evolution (\cite{Mitchell1996}). It repeatedly modifies a population of solutions. At each iteration (each generation in evolution), the algorithm selects individuals at random from the current population to be parents and reproduce offspring for the next generation, while allowing mutation and possibly elitism. Over successive generations, the population ``evolves" toward an optimal solution. 
The algorithm uses three rules to determine the reproduction for the next generation, namely the selection rule, the crossover rule and the mutation rule. The mutation rule is defined similarly as in Table~\ref{table:partition_neighbour_example}, whereas the others are slightly modified for factor collapsing using graycode format. The details of seting up the algorithm for factor collapsing can be found in Appendix~\ref{app:GA}.

\section{Results: Sweden TP claims data set}
\label{sec:small_example_benchmark}

\begin{figure}[hbt]
	\centering
	\begin{minipage}{0.45\linewidth}
		\centering
		\includegraphics[width=5cm, height=6cm]{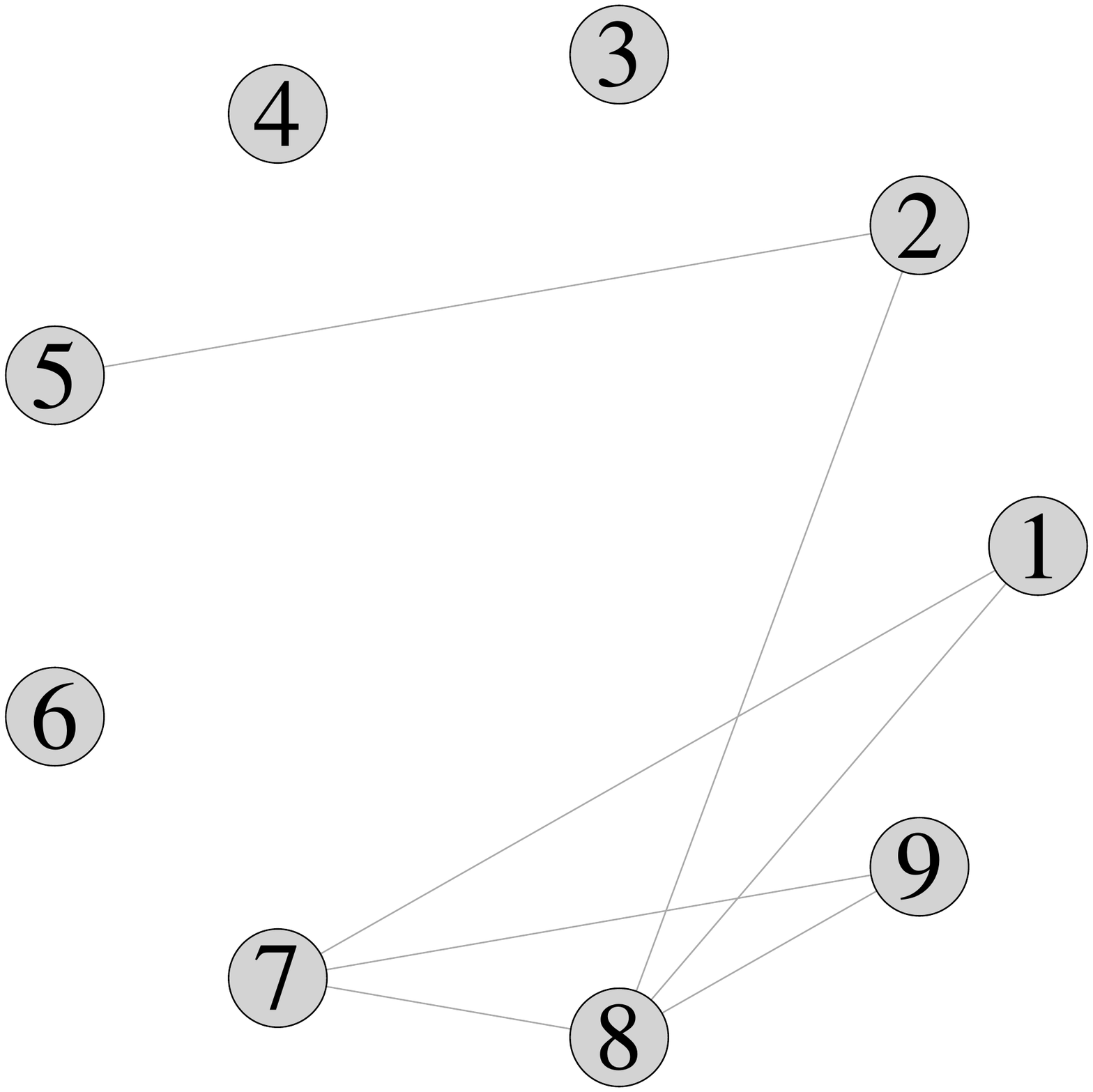}
		\caption*{(a)}
	\end{minipage}
	\begin{minipage}{0.45\linewidth}
		\centering
		\includegraphics[width=5cm, height=6cm]{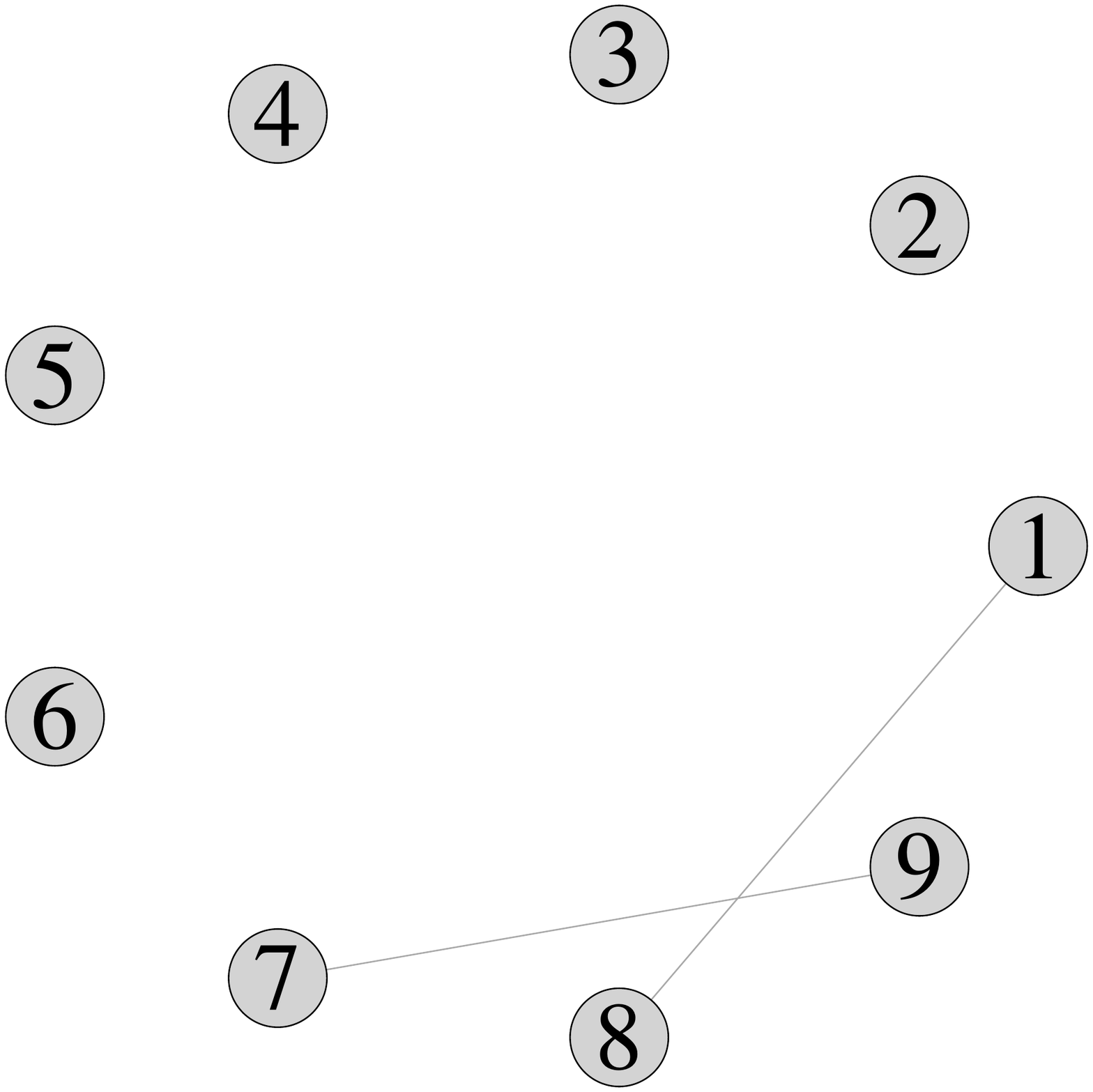}
		\caption*{(b)}
	\end{minipage}
	\captionof{figure}{Illustration of level equivalence for the rating factor Make: Figure~\ref{fig:graph_plot_make}(a) corresponds to the results of multiple comparisons in Table~\ref{table:multiple_comparison_make}; Figure~\ref{fig:graph_plot_make}(b) corresponds to the best result presented in Table~\ref{table:1var_factor_collapsing_no_interaction}. Each vertex represents one level and an edge between any two vertices means they are equivalent. }
	\label{fig:graph_plot_make}
\end{figure}

In this example, the 1977 Swedish TP motor insurance claim data is used. An over-dispersed Poisson GLM is first fitted for frequency, with claim count depending on the four variables, corrected for risk exposure. 
The model summary shows that all main effects are significant, but some levels within the rating factor ``Make" are not statistically significant and hence merging these levels seems natural in the next step. 
A first attempt could be merging the insignificant levels with the reference level but the best way to merge across all levels cannot be identified in this way. 
Traditionally a post-hoc analysis with multiple comparisons is implemented to compare the equivalence of means (i.e. levels) for the rating factor Make (\cite{Bondell2009}). The R package ``multcomp" is used (\cite{Hothorn2008simultaneous}). 
Table~\ref{table:multiple_comparison_make} is a subset of multiple comparison results on coefficients of Make, where only tested equivalent levels are shown.   
It shows that levels 1, 2, 5, 7, 8, 9 are statistically equivalent and these levels should be merged for model parsimony and homogeneity of risk, although equivalence between level 2 and 5 seems marginal and may imply uncertainty as to whether to cluster these two levels together or not. A clearer representation is shown in Figure~\ref{fig:graph_plot_make}. 
Therefore, we have four new levels: (125789)(3)(4)(6). 

%\begin{table}[ht]
%	\centering
%	\begin{minipage}{0.45\linewidth}
%		\caption{Multiple comparisons results on the factor Make in the frequency model: levels 1, 2, 5, 7, 8 and 9 are equivalent.}
%		\resizebox{\linewidth}{!}{
%		\begin{tabular}{c|ccccc}
%			\hline
%			Hypothesis & p-value \\
%			\hline    
%			coefficients of levels 7 and 9 equivalent &  0.9583  \\ 
%			coefficients of levels 8 and 9 equivalent &  0.4909   \\  
%			coefficients of levels 7 and 1 equivalent &  0.5600   \\  
%			coefficients of levels 8 and 1 equivalent &  1.0000   \\  
%			coefficients of levels 5 and 2 equivalent &  0.0839   \\
%			coefficients of levels 8 and 2 equivalent &  0.1429   \\  
%			coefficients of levels 8 and 7 equivalent &  0.9837  \\
%			\hline
%		\end{tabular} }
%		\label{table:multiple_comparison_make}    
%	\end{minipage}
%\begin{minipage}{0.47\linewidth}
%		\caption{Results for collapsing the factor Make in the frequency model. Here only the best 5 models (based on BIC) are shown.}
%	\label{table:1var_factor_collapsing_no_interaction}
%	\resizebox{\linewidth}{!}{
%	\begin{tabular}{ccccc}
%		\hline
%		\multicolumn{3}{l}{Make: 1, 2, 3, 4, 5, 6, 7, 8, 9} \\
%		\hline
%		Partition & BIC & Model weight \\
%		(1,8)(2)(3)(4)(5)(6)(7,9) &  10301.11 &  0.3458 \\  
%		(1,8)(2,5)(3)(4)(6)(7,9)  &  10301.81 &  0.2426 \\ 
%		(1,7,8)(2)(3)(4)(5)(6)(9) &  10303.44 &  0.1076  \\ 
%		(1,7,8)(2,5)(3)(4)(6)(9)  &  10304.15 &  0.0754  \\ 
%		(1)(2)(3)(4)(5)(6)(7,8,9) &  10304.92 &  0.0514 \\ 
%		\hline
%		\ \\
%	\end{tabular} }
%\end{minipage}
%	\end{table}

\begin{table}[ht]
	\centering
		\caption{Multiple comparisons results on the factor Make in the frequency model: levels 1, 2, 5, 7, 8 and 9 are equivalent.}
			\begin{tabular}{c|ccccc}
				\hline
				Hypothesis & p-value \\
				\hline    
				coefficients of levels 7 and 9 equivalent &  0.9583  \\ 
				coefficients of levels 8 and 9 equivalent &  0.4909   \\  
				coefficients of levels 7 and 1 equivalent &  0.5600   \\  
				coefficients of levels 8 and 1 equivalent &  1.0000   \\  
				coefficients of levels 5 and 2 equivalent &  0.0839   \\
				coefficients of levels 8 and 2 equivalent &  0.1429   \\  
				coefficients of levels 8 and 7 equivalent &  0.9837  \\
				\hline
		\end{tabular}
		\label{table:multiple_comparison_make}    
\end{table}

\begin{table}[ht]
	\centering
		\caption{Results for collapsing the factor Make in the frequency model. Here only the best 5 models (based on BIC) are shown.}
		\label{table:1var_factor_collapsing_no_interaction}
			\begin{tabular}{ccccc}
				\hline
				\multicolumn{3}{l}{Make: 1, 2, 3, 4, 5, 6, 7, 8, 9} \\
				\hline
				Partition & BIC & Model weight \\
				(1,8)(2)(3)(4)(5)(6)(7,9) &  10301.11 &  0.3458 \\  
				(1,8)(2,5)(3)(4)(6)(7,9)  &  10301.81 &  0.2426 \\ 
				(1,7,8)(2)(3)(4)(5)(6)(9) &  10303.44 &  0.1076  \\ 
				(1,7,8)(2,5)(3)(4)(6)(9)  &  10304.15 &  0.0754  \\ 
				(1)(2)(3)(4)(5)(6)(7,8,9) &  10304.92 &  0.0514 \\ 
				\hline
		\end{tabular}
\end{table}

By comparison, the FC method is run over the rating factor Make, keeping everything else unchanged in the model. There are $B_9 = 21,147$ models in the model space and an exhaustive search is performed to examine every model. The results are presented in Table~\ref{table:1var_factor_collapsing_no_interaction} and Figure~\ref{fig:graph_plot_make}. 
This result is slightly different from the result in Table~\ref{table:multiple_comparison_make}. In the best grouping, only levels 1 and 8, levels 7 and 9 are grouped together, but levels 2 and 5 are separated. In the second-best grouping, levels 2 and 5 are grouped together instead, with model weight being slightly lower than for the best model. Therefore the grouping is more granular using FC. BMA takes care of the uncertainty surrounding the grouping of levels 2 and 5 mentioned above.

Next the severity model is considered and a Gamma GLM is fitted, with average claim amount depending on the 4 predictive factors, corrected for exposure measured by number of claims. 
From the standard model summary, the factor Kilometers is not significant for predicting the claim amount. Hence a standard approach for the next step would be to eliminate this factor from the model entirely. 
If Kilometers is included in the model originally and the factor collapsing method is implemented, it is expected that this factor would have optimal collapsing as (12345), i.e. all levels combined.
The model summary also shows that for the factor Make, levels 2 and 6 are not significant and hence should be merged with the reference level. 

When running FC over the factor Kilometers the best collapsing result is (1)(23)(45) instead of (12345) as anticipated (Table~\ref{table:1var_factor_collapsing_severity}).
This means that even though by standard analysis Kilometers should be excluded in variable selection, merging some levels together leads to this factor being significant. 
It also implies that when using FC-BMA, a relatively large (or saturated) model should be used as a baseline model before FC in case some predictive factors become significant after collapsing.
It is also noted that while Kilometers is treated as nominal, the best results preserve an increasing order, indicating that treating this factor as ordinal is a reasonable alternative.

As a comparison to the frequency model above, Make is also collapsed in the severity model, as is shown in Table~\ref{table:1var_factor_collapsing_severity}. It turns out that the optimal result is to combine levels 2 and 6 together while keeping all other levels unchanged with probability close to one. Therefore, although levels 2 and 6 are not significant by themselves, grouping them makes the new level significant.  
One would expect that variables are of differing significance in the frequency model and severity model separately (\cite{Frees2016}).
%When ``Kilometers” is deleted (or merging all its levels), the BIC value is 1874800 (with log likelihood as -937182.8 (df=58)).
%That means factor collapsing method will never pick up the case when ``Kilometers” should be excluded, as suggested by the standard model.
%If we keep the ``Kilometers" in the model and then run our method, the best BIC of 1874300 is less than that of the standard model.

\begin{table}[H]
	\centering
	\caption{Results for collapsing Make and Kilometers individually in the severity model. In both cases only the best 5 models (based on BIC values) are shown.}
	\label{table:1var_factor_collapsing_severity}
	\resizebox{\textwidth}{!}{
	\begin{tabular}{lcc||lcc}
		\hline
		\multicolumn{3}{l||}{Make: 1, 2, 3, 4, 5, 6, 7, 8, 9} & 			\multicolumn{3}{l}{Kilometers: 1, 2, 3, 4, 5} \\
		\hline
		Partitions & BIC & BMA weight & Partition & BIC & BMA weight \\
		(1)(26)(3)(4)(5)(7)(8)(9)   & 1,878,164 &  0.9673  & (1)(23)(45)    &1,878,161 &  0.8124 \\
		(1)(2)(3)(4)(5)(6)(7)(8)(9) & 1,878,171 &  0.0327 & (1)(2)(3)(45)   &1,878,164 &  0.1430 \\
		(1)(29)(3)(4)(5)(6)(7)(8)   & 1,878,189 &  0.0000 & (1)(23)(4)(5)   &1,878,167 &  0.0379 \\
		(1)(26)(3)(4)(57)(8)(9)     & 1,878,190 &  0.0000 & (1)(2)(3)(4)(5) &1,878,170 &  0.0067 \\
		(1)(2)(3)(4)(5)(69)(7)(8)   & 1,878,190 &  0.0000 & (1)(25)(3)(4)   &1,878,198 &  0.0000 \\
		\hline
	\end{tabular}  }  
\end{table}

Continuing the example when collapsing the four factors Kilometers, Zone, Bonus and Make simultaneously, the Bell numbers for the four factors are $B_5 = 52, B_7 = 877, B_7 = 877$ and $B_9 = 21147$ respectively, therefore there are $B_5*B_7*B_7*B_9 \approx 8.4576*10^{11}$ models in the model space that need to be searched across, which is very computationally intensive.
Hence stochastic optimisation is used to collapse all four rating factors simultaneously using both SA and GA. To assess stability of the results, the algorithms have been run 20 times.
Table~\ref{table:4var_factor_collapsing_frequency} and Talbe~\ref{table:4var_factor_collapsing_severity} give the results for both the frequency model and the severity model.

\begin{table}[htb]
	\centering
	\caption{Frequency model results for collapsing Kilometers, Zone, Bonus and Make simultaneously. The best 5 models based on BIC are shown. They account for 89\% posterior probability between them. }
	\resizebox{\textwidth}{!}{
	\begin{tabular}{llllcccc}	
		\hline
		Kilometers & Zone & Bonus & Make & BIC  & BMA weights\\   
		\hline
		(1)(2)(3)(4)(5)  & (1)(2)(3)(4,7)(5)(6)  & (1)(2)(3)(4)(5)(6)(7)  & (1,8)(2)(3)(4)(5)(6)(7,9)  & 10297.2  & 0.3765    \\
		(1)(2)(3)(4)(5)  & (1)(2)(3)(4,7)(5)(6)  & (1)(2)(3)(4)(5)(6)(7)  & (1,8)(2,5)(3)(4)(6)(7,9)   & 10298.0  & 0.2578    \\
		(1)(2)(3)(4)(5)  & (1)(2)(3)(4,7)(5)(6)  & (1)(2)(3)(4)(5)(6)(7)  & (1,7,8)(2)(3)(4)(5)(6)(9)  & 10299.5  & 0.1212  \\
		(1)(2)(3)(4)(5)  & (1)(2)(3)(4,7)(5)(6)  & (1)(2)(3)(4)(5)(6)(7)  &	(1,7,8)(2,5)(3)(4)(6)(9)   & 10300.2  & 0.0829    \\
		(1)(2)(3)(4)(5)  & (1)(2)(3)(4,7)(5)(6)  & (1)(2)(3)(4)(5)(6)(7)  & (1)(2)(3)(4)(5)(6)(7,8,9)  & 10301.1  & 0.0541  \\
		\hline
	\end{tabular} }
	\label{table:4var_factor_collapsing_frequency}
\end{table}

\begin{table}[htb]
	\centering
	\caption{Severity model results for collapsing Kilometers, Zone, Bonus and Make simultaneously. The best 5 models based on BIC are shown. They account for 95\% posterior probability between them.}
	\resizebox{\textwidth}{!}{
	\begin{tabular}{llllcccc}	
		\hline
		Kilometers & Zone & Bonus & Make & BIC  & BMA weights\\   
		\hline
		(1)(2,3)(4,5)  & (1)(2,7)(3,5)(4)(6)  & (1)(2)(3,6)(4)(5)(7)  &  (1)(2,6)(3)(4)(5)(7)(8)(9) & 1878133.8 & 0.6552 \\
		(1)(2,3)(4,5)  & (1)(2,7)(3,5)(4)(6)  & (1)(2,5)(3,6)(4)(7)  & (1)(2,6)(3)(4)(5)(7)(8)(9)  & 1878137.1 & 0.1296 \\
		(1)(2)(3)(4,5)  & (1)(2,7)(3,5)(4)(6)  & (1)(2)(3,6)(4)(5)(7)  & (1)(2,6)(3)(4)(5)(7)(8)(9)  & 1878137.4 & 0.1120 \\
		(1)(2)(3)(4,5)  & (1)(2,7)(3,5)(4)(6)  & (1)(2,5)(3,6)(4)(7)  & (1)(2,6)(3)(4)(5)(7)(8)(9)  & 1878140.1 & 0.0287 \\
		(1)(2,3)(4,5)  & (1)(2,7)(3,5)(4)(6)  & (1)(2)(3,6)(4)(5)(7)  & (1)(2)(3)(4)(5)(6)(7)(8)(9) & 1878140.7 & 0.0212 \\
		\hline
	\end{tabular} }
	\label{table:4var_factor_collapsing_severity}
\end{table}

To verify the FC-BMA method indeed works better for prediction, Table~\ref{table:example_prediction_comparison_frequency} and Table~\ref{table:example_prediction_comparison_severity} show the results of FC-BMA, as well as some other methods in the literature previously mentioned in the Introduction section of this article that can be interpreted as factor collapsing methods. The details for these methods can be found in Appendix~\ref{app:benchmark}.
The data set is divided into training and test sets (80\% and 20\% respectively) and out of sample predictions are calculated using various metrics: Gini index (\cite{Frees2014gini}), concordance correlation coefficient (\cite{Cox2006}), Wasserstein distance (\cite{Vallender1974}), Kolmogorov-Smirnov test (\cite{Boland2007}), KL divergence (\cite{Kullback1951}), root mean squared error (\cite{Lehmann2006}). Note that the results are averages of 50 repetitions to ensure result stability when randomly splitting the data. 
In Table~\ref{table:example_prediction_comparison_frequency} either FC-only or FC-BMA shows better prediction accuracy using most metrics except Kolmogorov-Smirnov test. The improvement found using BMA varies, but in most cases it improves the prediction. It is also worth noting that even though in some cases the improvement is only slight versus the standard GLM, the model complexity is much reduced.  
Similarly in Table~\ref{table:example_prediction_comparison_severity}, either FC-only or FC-BMA show the most satisfactory results depending on which metric is chosen.

\begin{table}[htb]
	\centering
	\caption{Number of claims (frequency) prediction comparison when splitting the data set into 80\% training and 20\% test data, using Gini index, concordance correlation coefficient (CCC), Wasserstein distance, Kolmogorov-Smirnov test (KS-test), KL divergence and root mean squared error (rMSE) respectively. The underlined values are the best models according to each metric.}
	\begin{tabular}{lcccccc}
		\hline
		\  & Gini & CCC  & Wass. & KS-test & KL & rMSE \\
		\hline
		no FC-BMA  & 0.8266 & 0.9968 & 3.0340 & 0.0736(0.30) & 0.0122& 16.3383 \\
		FC-only    & 0.8267 & 0.9943 & \underline{2.9696} & 0.0788(0.24) & 0.0114& \underline{14.9927} \\
		FC-BMA(5)  & \underline{0.8267}	& \underline{0.9973} & 4.2012 & 0.0778(0.25) & \underline{0.0113}& 21.3630 \\
		Regression Tree &	0.8246 	&  0.9732  & 6.4821    & \underline{0.0694(0.35)} &0.0450 & 41.3327  \\
		Multiple comparison& 0.8202 &0.9651 &6.8543    & 0.0972(0.07)  & 0.0313 & 49.3326 \\
		BMA R package  &0.7845 	&0.9921	 & 3.2907 & 0.0806(0.19) & 0.0196  & 16.2055 \\
		\hline
	\end{tabular}
	\label{table:example_prediction_comparison_frequency}
\end{table}

\begin{table}[htb]
	\centering
	\caption{Average claim amount (severity) prediction comparison when splitting the data set into 80\% training and 20\% test data, using Gini index, concordance correlation coefficient (CCC), Wasserstein distance, Kolmogorov-Smirnov test (KS-test), KL divergence and root mean squared error (rMSE) respectively. The underlined values are the best models according to each metric.}
	\begin{tabular}{lcccccc}
		\hline
		\  & Gini & CCC  & Wass. & KS-test & KL & rMSE \\
		\hline
		no FC-BMA & 0.0567 & 0.0409 & 1948.3340 & 0.4489(0) & 0.2191 & 3840.3717 \\
		FC-only   & 0.0576 & \underline{0.0667} & 1825.0540 & 0.4067(0) & 0.2178 & \underline{3829.4343} \\
		FC-BMA(5) &  0.0576 & 0.0657 & \underline{1822.9450} & \underline{0.4033(0)} & \underline{0.2178}& 3829.6677 \\
		Regression Tree     & 0.0512 & 0.0262  & 2090.8150 & 0.5333(0) & 0.2562 & 4169.6031 \\
		Multiple comparison&0.0111  & 0.0000	 & 2921.8380 & 0.5694(0) & 0.2771 & 4893.2215  \\
		BMA R package   &\underline{0.0897}  &0.0459 & 2280.6890 & 0.4583(0) &  0.2824  & 4858.6437  \\
		\hline
	\end{tabular}
	\label{table:example_prediction_comparison_severity}
\end{table}

\section{Results: Irish motor insurance data}
\label{sec:real_example}

This section illustrates a complete case study for implementing the FC-BMA method using the Irish insurer accidental damage claim data introduced in Section~\ref{sec:data}.
Frequency and severity baseline models are built, using only main effects of the selected variables, part of which are shown in Table~\ref{table:case_study_dataset_pickedvar}.
Among the selected variables, there are 19 rating factors each with less than 15 categorical levels and four factors with more than 30 categories each. %and 4 variables are treated as continuous. 
Considering computational complexity when all factors are collapsed simutaneously, it is decided to collapse the first 19 factors in unison, then the four variables with very high numbers of levels individually. Then, the best few selected partitions in each case are combined again to search for the best combinations among them.  
%First, 19 rating factors each of which has less than 15 levels are collapsed simultaneously. 
%Next, rating factors who have substantial number of levels including "County", "Mosaic Type", "ABI vehicle rating group", "M.D. licence length" are collapsed individually, it is because it would be too computationally intensive to collapse all of them simultaneously. 
Computation has been repeated 10 times to verify the stability of the results. In this example, the behaviour of the county factor is focused on, where there are 35 levels, including all 26 counties (Tipperary is treated as South Tipperary and North Tipperary), 4 local authorities in county Dublin (Dublin City, Dun Laoghaire-Rathdown, Fingal, South Dublin), 4 other major cities in Ireland (Cork, Galway, Limerick, Waterford) and Unknown. For data sensitivity issues only part of the 19 factors result is shown with corresponding model coefficients. Results for prediction accuracy and comparison are shown in Table~\ref{table:case_study_prediction_comparison}. 

\begin{figure}[H]
	\centering
	\begin{minipage}{0.45\textwidth}
		\includegraphics[width=6cm, height=7.5cm]{FreqModCountyCoef_new.pdf}
		\caption*{(a) Frequency}
	\end{minipage}
	\begin{minipage}{0.45\textwidth}
		\includegraphics[width=6cm, height=7.5cm]{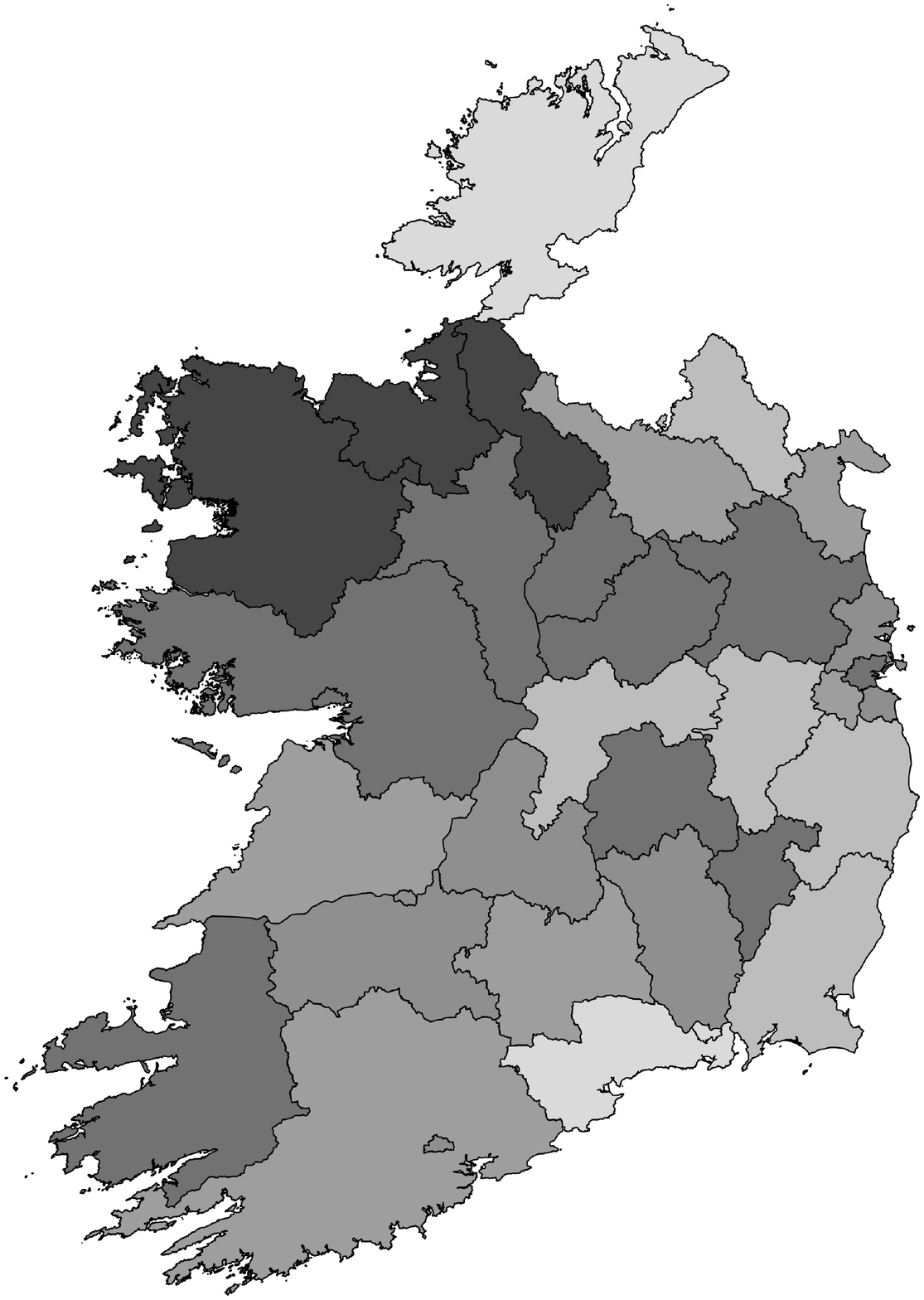}
		\caption*{(b) Severity}
	\end{minipage}	
	\caption{Map of Irish counties illustrating the best collapsing result for the frequency model: in Fig~\ref{fig:case_study_frequency_best_collapsing}(a) colors correspond to estimated coefficients based on a standard GLM fit from Table~\ref{table:county_coef_frequency}. In Fig~\ref{fig:case_study_frequency_best_collapsing}(b) colors correspond to estimated coefficients based on the best factor collapsing result (highest model weight). The darker the color the higher the frequency of making a claim in the county in question. }
	\label{fig:case_study_frequency_best_collapsing}
\end{figure}

\begin{figure}[H]
	\centering
	\begin{minipage}{0.45\textwidth}
		\includegraphics[width=6cm, height=7.5cm]{SevModCountyCoef_new.pdf}
		\caption*{(a) Frequency}
	\end{minipage}
	\begin{minipage}{0.45\textwidth}
		\includegraphics[width=6cm, height=7.5cm]{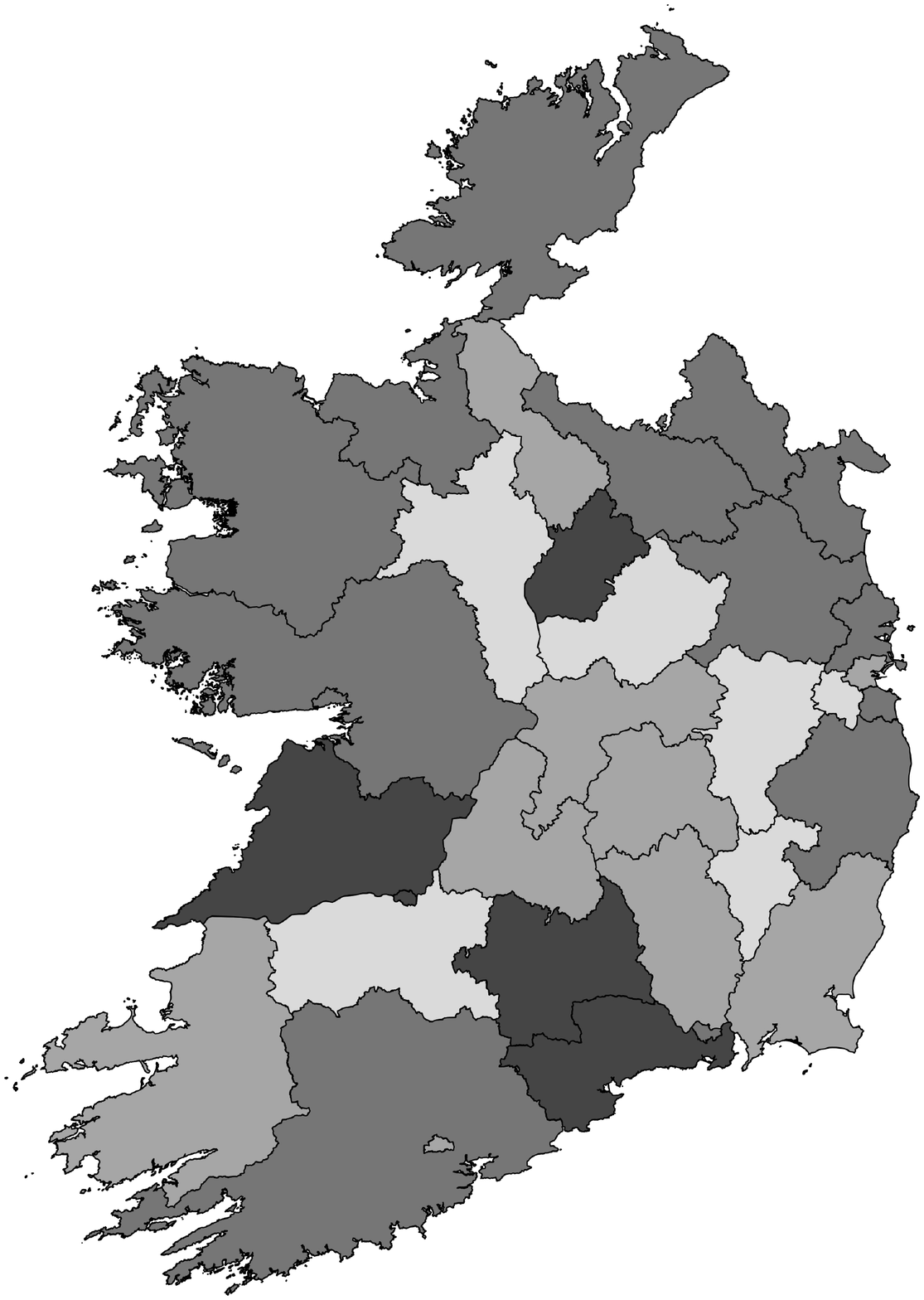}
		\caption*{(b) Severity}
	\end{minipage}
	\caption{Map of Irish counties illustrating the best collapsing result for the severity model: in Fig~\ref{fig:case_study_severity_best_collapsing}(a) colors correspond to estimated coefficients based on a standard GLM fit from Table~\ref{table:county_coef_severity}. In Fig~\ref{fig:case_study_severity_best_collapsing}(b) colors correspond to estimated coefficients based on the best factor collapsing result (highest model weight). The darker the color the higher the claim amount when a claim is made in the county in question.}
	\label{fig:case_study_severity_best_collapsing}	
\end{figure}

Baseline models are first fitted for frequency and severity respectively, where in the frequency model claim count is regressed against all predictors, corrected for risk exposure. In the severity model average claim cost is regressed against all predictors, corrected for number of claims. 
There is no clear inherent ordering in county categories. Therefore it is treated as nominal and adjacent counties do not have to be of similar risk profiles.
The model coefficients for counties in the baseline models are listed in Table~\ref{table:county_coef_frequency} and Table~\ref{table:county_coef_severity} in increasing order. All coefficients are very close to the adjacent coefficient values and some of them are not significantly different from the reference level (Carlow county). Figure~\ref{fig:case_study_frequency_best_collapsing} and Figure~\ref{fig:case_study_severity_best_collapsing} show the coefficients for each county on an Irish map.

By repeatedly implementing the stochastic search algorithm, it has been shown that the algorithm is stable sufficiently to always find the global minimum BIC value. For the frequency part, there are six clusters of counties, each cluster containing counties having adjacent values of coefficients obtained from the standard GLM, as shown in Table~\ref{table:county_coef_frequency}. Note that the clustering shown in Table~\ref{table:county_coef_frequency} comprises only the five best FC results, which account for 11.24\% posterior model weight. The reason for choosing five models is because the first 50 models account for 80\% model probability and the first 149 models account for 100\% model probability, thus it is impractical to show coefficients of all of them. In particular, the top few clusterings are very similar, except only one category changes from one group to another, mostly around the boundaries between clusters. Figure~\ref{fig:case_study_frequency_best_collapsing}(b) shows the clustering results on an Irish map. Most clusters consist of geographically adjacent counties, such as counties in the north-west part of Ireland including Sligo, Leitrim and Mayo as having the highest probability of making a claim. Counties on and adjacent to the east coast including Kildare, Wicklow and Wexford have the lowest probabilities of making a claim. It also reveals some interesting insights, for example Donegal and Waterford, while being far away from each other and on the opposite sides of Ireland, have very similar risk profile in terms of claim frequency. 

In the severity part, similar patterns are shown in Table~\ref{table:county_coef_severity}, where in most cases, particularly in the best collapsing, only adjacent levels from the standard GLM are combined, except the level Unknown is changing across different clusterings. Some interesting insights are also found in Figure ~\ref{fig:case_study_severity_best_collapsing}(b). There are four clusters found and most of them have mainly geographically adjacent counties grouped together. One interesting observation is, for counties like Waterford and Tiperrary, while they have relatively low probability of making a claim in frequency terms, once there is a claim made the severity is usually high. 

\begin{landscape}
	\begin{figure}[H]
		\centering
		\begin{minipage}{0.7\textwidth}
			\centering
			\includegraphics[width=\linewidth, height=1.1\linewidth]{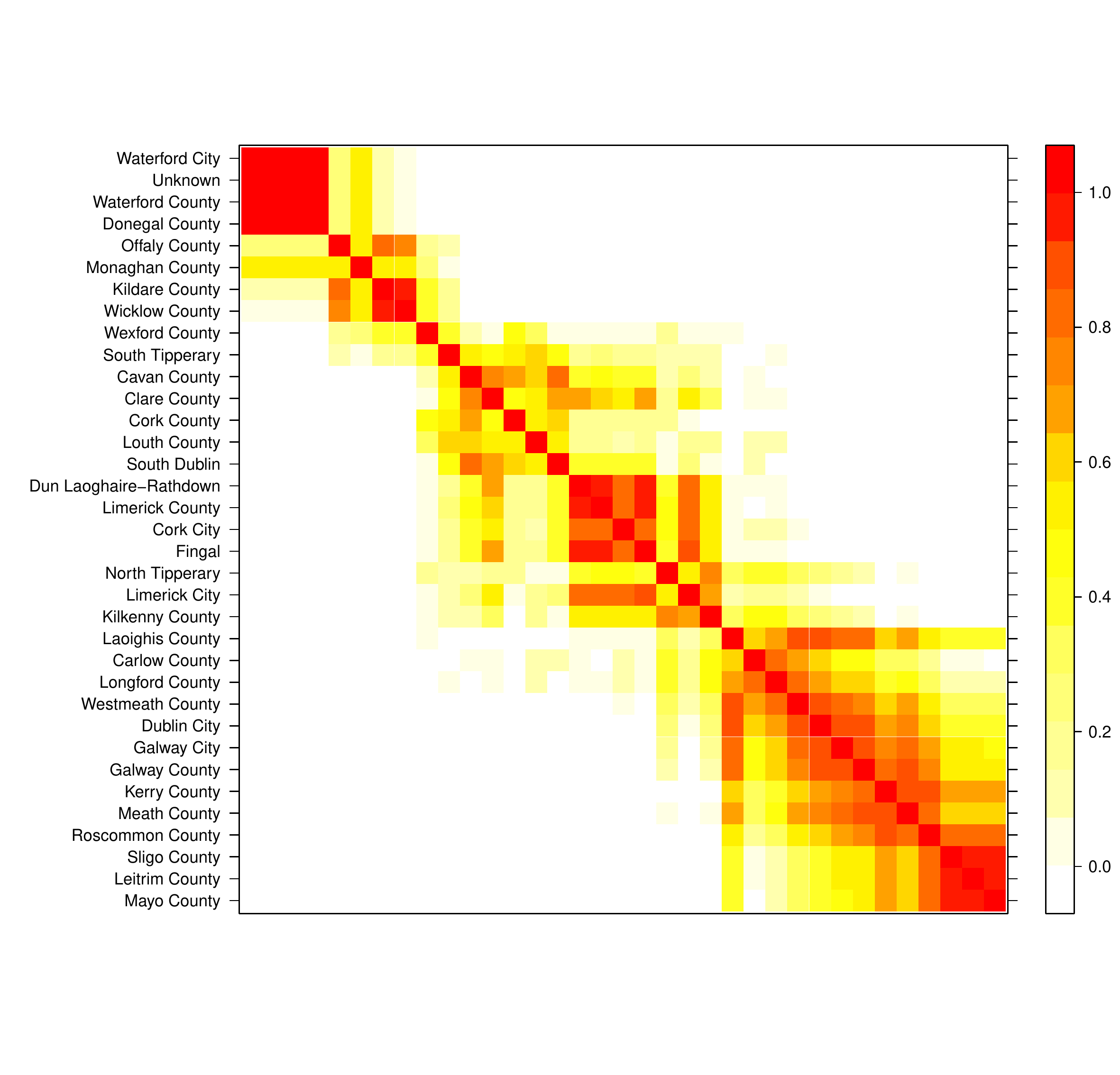}
			\caption*{(a) Frequency}
		\end{minipage}
	\hspace{1cm}
		\begin{minipage}{0.7\textwidth}
			\centering
			\includegraphics[width=\linewidth, height=1.1\linewidth]{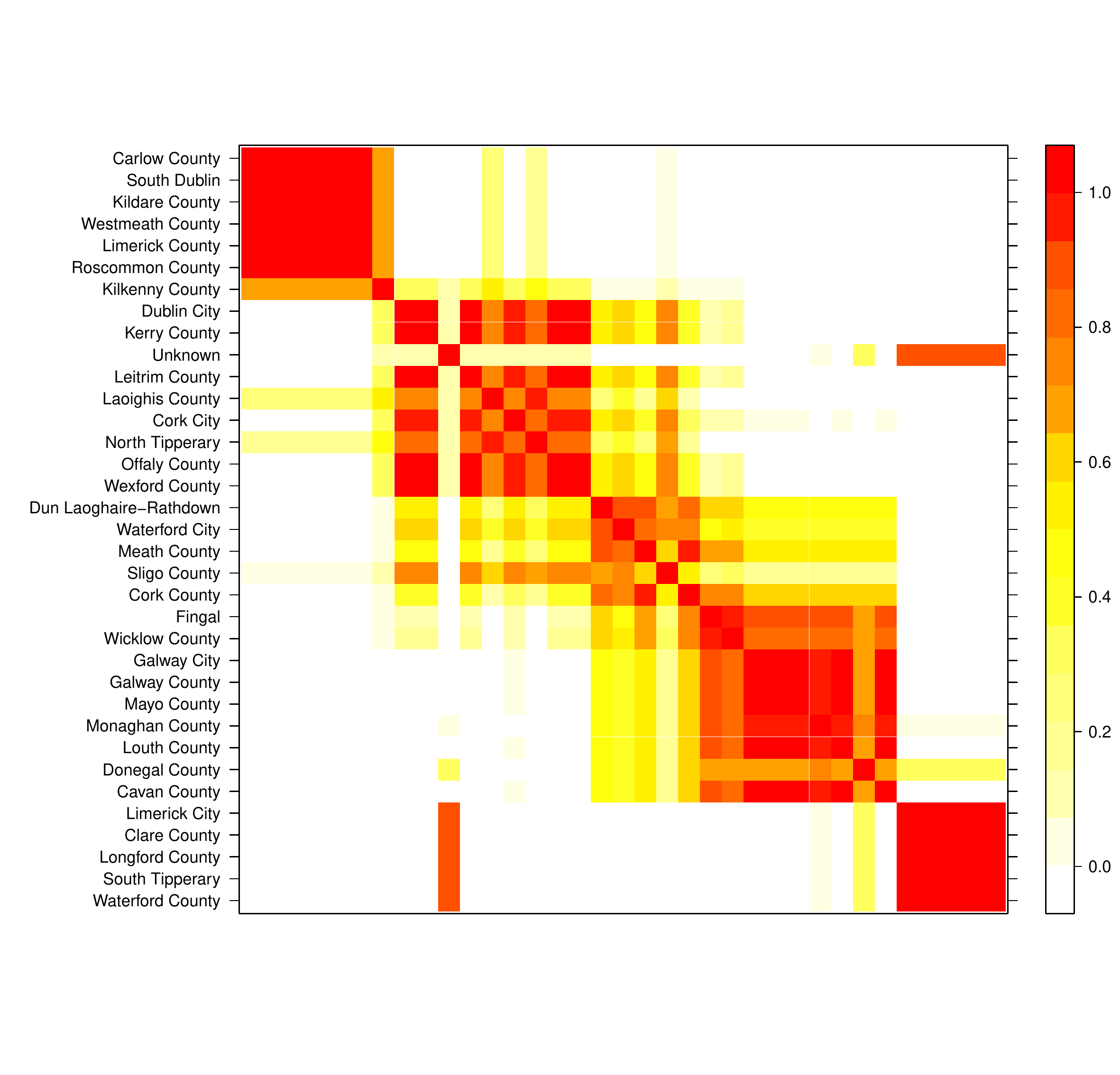}
			\caption*{(b) Severity}
		\end{minipage}
		\caption{Clustering results using FC-BMA for the county variable using a binary similarity matrix: (a) the frequency part corresponding to the best 66 models selected; (b) the severity part correponding to the best 17 models selected. The darker the color, the higher the probability that two levels belong to the same group.}
		\label{fig:clustering_heatmap}
	\end{figure}
\end{landscape}

Figure~\ref{fig:clustering_heatmap} shows a heatmap for clustering among categorical levels, illustrating posterior probabilities of any pair of categories being in the same cluster. The darker the color, the higher the probability of the pair being in the same cluster among all selected models, corresponding to Table~\ref{table:county_coef_frequency} and Table~\ref{table:county_coef_severity}. In the plots, the order of counties are again based on the coefficients of the standard GLMs. It shows the differences among all selected partitions are mostly due to changes of the levels at the boundaries of clusters. 
It is noted that although the county variable is not necessarily a good predictor for claims because it is a less granular geo-variable, its having 35 categories demonstrates the algorithm is efficient in stochastic optimisation. Our experience shows that when the number of categories is approximately 50 or higher for one factor, FC-BMA works relatively well. One should be more cautious in deciding how many factors can be collapsed simultaneously based on the numbers of categorical levels.  

Next, 19 factors are collapsed simultaneously and part of the best FC result is shown in Table~\ref{table:case_study_19_factors} for frequency and severity respectively. Because of data security, not all collapsed factors are shown and categories in some factors are masked. For frequency, out of all 16 factors, 10 factors turn out to be significant; whereas for severity only 4 out of 16 factors are significant. This is expected as it is acknowledged that in general the frequency model requires more predictors than the severity model (\cite{Charpentier2014}; \cite{Coutts1984} and \cite{Frees2014}). 
For both parts, the best FC result shown only accounts for less than 10\% posterior model probability and among the best FC combinations there is only one element altered, meaning their BIC values are very close. %For frequency part, the top 20 models account for about 80\% posterior model weights, similarly for severity. 
%Other categorical variables involved in the model such as vehicle categories and main driver licence length are also collapsed but will not be discussed in detail here. 

\begin{table}[H]
	\centering
	\caption{Selected model coefficients of the best collapsed GLM for frequency (left) and severity (right) when simultaneously collapsing 19 factors. Not all variables are shown. The variables detailed correspond to those in Table~\ref{table:case_study_dataset_pickedvar}. The symbol $\ast$ represents the reference level in GLM fitting using reference level dummy coding. If all the categorical levels within a predictor are collapsed together with $\ast$ as a coefficient, it means that this predictor is excluded from the model.}
	\label{table:case_study_19_factors}
\begin{minipage}{0.48\linewidth}
	\resizebox{\textwidth}{!}{
		\begin{tabular}{llc}
			\hline
			Predictors& Categorical levels & Coefficients \\
			\hline
			Intercept & \ & -17.2029 \\
			Policyholder gender& Female; Male; Neutral & $\ast$  \ \\
			Penalty point& 0; 7-8 & $\ast$ \\ 
			\            & 1-6, 9+ & -0.1263 \\
			Vehicle fuel type& Diesel, Petrol & $\ast$ \\
			\        & Unknown& 0.7197 \\
			Vehicle transmission& Automatic; Manual; Unknown & $\ast$ \\
			Annual mileage & 0-15000; 25001-50000 & $\ast$  \\
			\ &       15001-25000; 50001+ & 0.2341 \\                         
			%Driver option& (INSDONLY, INSDSPSE, NAMEDDRV, OPENDRIV)& \ \\
%			Class of use& (SDP only)& \ \\ 
%			\           & (SDP/retail/publican, SDP/trade(inc empl), SDP/trade(inc empl,agnts), SDP/trade(insd only))& \ \\
			Number of registered drivers & 1; 2; 3; 4; 5; 6; 7 & $\ast$  \\
			No claim discount (NCD) & 0   & $\ast$ \\ 
			\   & 0.1 & -2.2386 \\ 
			\   & 0.2 & -1.1407 \\ 
			\   & 0.3; 0.4& 12.7243  \\ 
			\   & 0.5; 0.6& 10.2648 \\
			NCD protection&  No; Unknown & $\ast$ \\
			\    & Yes & -12.7728  \\
			NCD stepback &  No & $\ast$  \\
			\            &  Yes& 0.1569  \\
			Total excess (\euro)& 0; 300; 600 & $\ast$ \\
			\           & 125 & 0.3775  \\
			Main driver licence category& B; D; I; N & $\ast$  \\
			\               & C; F & 0.0899  \\
			%Second car & Higher cc; Lower cc & $\ast$ \\
			%\          & No second car & 0.2250  \\
			%Product & (An Post direct, BrokerNET, Motorcare, Motorchoice)& \  \\  
			%Pay plan & No & $\ast$ \\ 
			%\        & Yes; Yes with 0 charge &  -0.0464 \\
			%Licence FP & (full+0prov, full+1prov, full+2+prov, prov+0prov, prov+1+prov)& \ \\        
%			SRC channel &(BANCA-AIB, BANCA-AN POST, BANCA-FORD, DIRECT-BRANCH, DIRECT-TELESALE)& \ \\
%			\ &(BANCA-ULSTER, BROKER, BROKER-BFI, DIRECT-CCWEB, DIRECT-INTERNET)& \  \\
%			\ &(DIRECT-STAFF)         & \   \\
%			Ignition discount &(0\%, 30\%, P)& \  \\
%			\ &(10\%, 10\% Curfew Only, 20\%, 40\%, 40\% with curfew, IAM 20\%)& \ \\
%			\ &(20\% with curfew, 30\% with curfew)  & \   \\
			\hline
	\end{tabular}	}
\end{minipage}
\begin{minipage}{0.48\linewidth}
	\resizebox{\textwidth}{!}{
	\begin{tabular}{llc}
		\hline
		Predictors& Categorical levels & Coefficients \\
		\hline
		Intercept& \ & 9.6906 \\ 
		Policyholder gender& Female; Male; Neutral& $\ast$  \\
		Penalty point& 0, 1-2, 3-4, 5-6, 7-8, 9+ & $\ast$\\
		Vehicle fuel type& Diesel; Petrol; Unknown& $\ast$ \\
		Vehicle transmission& Automatic; Manual; Unknown & $\ast$ \\
		%Driver option& (INSDONLY, INSDSPSE, NAMEDDRV, OPENDRIV)& \ \\
		%Class of use&(SDP only, SDP/retail/publican, SDP/trade(inc empl) SDP/trade(inc empl,agnts) SDP/trade(insd only)) & \  \\
		Annual mileage      & 0-20000; 25001-40000 & $\ast$ \\
		\ & 20001-25000; 40001+  & 0.0772    \\
				Number of registered drivers&  1; 2; 5 & $\ast$ \\
		\            &  3; 4; 6; 7 & 0.1367 \\
		No claim discount (NCD)&  0; 0.1; 0.2; 0.3; 0.4; 0.5; 0.6 & $\ast$ \\
		NCD protection&   N & $\ast$  \\ 
		\             &   Unknown; Y & -0.2960 \\
		NCD stepback&  N; Y  & $\ast$ \\
		Total excess (\euro)&  0; 125; 300; 600 & $\ast$ \\
		Main driver licence group&  B; F; I; N & $\ast$ \\ 
		\               & C; D & 0.0952 \\ 
		%Second car& Higher cc; Lower cc; No second car & $\ast$ \\
		%Product& (An Post direct, BrokerNET, Motorcare, Motorchoice)& \ \\
		%Pay plan&  N; Y; Y with 0 charge & $\ast$ \\
		%Licence FP&(full+0prov, prov+0prov, prov+1+prov)& \  \\ 
		%\        & (full+1prov, full+2+prov)& \ \\  
		%			SRC channel &(BANCA-AIB, BANCA-AN POST, BANCA-FORD, BROKER-BFI, DIRECT-BRANCH, DIRECT-TELESALE)& \ \\
		%			\ &(BANCA-ULSTER, DIRECT-STAFF)   & \   \\
		%			\ &(BROKER, DIRECT-CCWEB, DIRECT-INTERNET) & \   \\
		%			Ignition discount & (0\%, 20\%, 20\% with curfew, 30\%, P)& \ \\
		%			\ &(10\%, 10\% Curfew Only)  & \    \\     
		%			\ &(30\% with curfew, 40\% with curfew, IAM 20\%)& \ \\
		%			\ &(40\%)& \  \\
		\hline
		\ \\
		\ \\
		\ \\
		\ \\
		\ \\
		\ \\
		\ \\
	\end{tabular}	}
\end{minipage}
\end{table}

After all factors are collapsed, predictions are compared to assess the efficiency and predictive power of FC-BMA. In Table~\ref{table:case_study_prediction_comparison}, predictions using the standard GLM, predictions using FC-only (the best FC result with highest model weight) and predictions using FC-BMA are compared. In the frequency model, it is as expected that in most cases FC-BMA gives the best predictions. When using concordance correlation coefficient or Kolmogorov-Smirnov test, FC-only gives the best predictions but FC-BMA is only marginally different. This confirms that BMA improvement varies based on choice of metric. 
In the severity part, FC-BMA gives the best prediction in four cases and FC-only is marginally worse. Other metrics give the best results for no-FC (i.e. the standard GLM). In general, results for severity are not as satisfactory as those for frequency. Considering Table~\ref{table:case_study_19_factors} this makes sense, because out of 19 factors in the severity model, only 7 are statistically significant (although in Table~\ref{table:case_study_19_factors} only 4 significant predictors are shown). Thus, using FC-only or FC-BMA reduces much of the model complexity. In comparison, the standard GLM using all predictors is closer to the saturated model and while it gives slightly better prediction, it lacks model parsimony.

\begin{table}[H]
	\centering
	\caption{Prediction accuracy comparison when splitting the full data set into 80\% training data and 20\% test data, using Gini index, concordance correlation coefficient (CCC), Wasserstein distance (Wass.), Kolmogorov-Smirnov test (KS-test), Kullback-Leibler divergence (KL) and root mean squared error (rMSE). In the frequency segment 6000 best models (i.e. collapsings) are averaged, while in the severity segment 3344 models are averaged. }
	\label{table:case_study_prediction_comparison}
	\resizebox{\textwidth}{!}{
	\begin{tabular}{llrrrrrrrrr}
		\hline
		\ & \ & Gini Index  & CCC & Wass.  & KS-test & KL & MSE \\
		\hline
		\multirow{3}{*}{Frequency} & no FC   & 0.7000  & 0.0489 & 0.0347  & \underline{0.9800 (0)} & 3.6127& 0.1428 \\
		& FC only & 0.7016  & \underline{0.1078} & 0.0337 & \underline{0.9800 (0)} & 3.5122 & 0.1404 \\
		& FC-BMA(6000)  & \underline{0.7019}  & 0.0977 & \underline{0.0335}  & 0.9806 (0) & \underline{3.5013}& \underline{0.1378} \\
		\hline
		\multirow{3}{*}{Severity} & no FC   & 0.5565  & 0.0559 & \underline{575.2057}  & \underline{0.2141 (0)} & 0.4573& 4017.6134 \\
		& FC only & 0.5745  & \underline{0.1602} & 855.8074  & 0.3264 (0) &  0.3328& 2108.5297 \\
		& FC-BMA(3344)  & \underline{0.5747}  & \underline{0.1602} & 858.6310  & 0.3242 (0) &  \underline{0.3323} & \underline{2106.9606} \\
		\hline
	\end{tabular} }
\end{table}

\begin{table}[htb]
	\centering
	\caption{The frequency model coefficients for the standard GLM and the GLMs using factor collapsing. Categorical levels are of increasing order based on the standard GLM coefficients for ease of illustration. Among all selected models, the first 50 models and the first 149 models account for 80\% and 100\% model posterior probability respectively. Only the five best models are selected for ease of illustration. Note that for the five models shown only adjacent categorical levels are clustered. But as model weight becomes smaller, this will occur less often.}
	\label{table:county_coef_frequency}
	\resizebox{\linewidth}{!}{
		\begin{tabular}{lrrrrrrrrrrrrrrrrrrrrrrrrrrrr}
			\hline\hline
			\ &	Standard GLM &Model 1&Model 2&Model 3&Model 4&Model 5\\%&Model 6&Model 7 \\
			\hline
			BIC& \ & \ 62,807.2927&\ 62,807.3039&	\ 62,807.3972& \ 62,807.4069&	\ 62,807.4294\\%&	62807.4471&	62807.5250\\
			Model weight among all models& \  &  0.0233&	0.0232&	0.0221&	0.0220&	0.0218\\%&	0.0216&	0.0208 \\
			\ \\
			Waterford City	&-6.6556&-6.6414&-6.6399&-6.6326&-6.6341&-6.6311\\%&-6.5955&-6.6306\\
			Unknown		  	&-6.6130&-6.6414&-6.6399&-6.6326&-6.6341&-6.6311\\%&-6.5955&-6.6306\\
			Waterford County&-6.6073&-6.6414&-6.6399&-6.6326&-6.6341&-6.6311\\%&-6.5955&-6.6306\\
			Donegal County 	&-6.5959&-6.6414&-6.6399&-6.6326&-6.6341&-6.6311\\%&-6.5955&-6.6306\\ %\cline{4-4}
			Offaly County 	&-6.5787&-6.5733&-6.6399&-6.6326&-6.6341&-6.6311\\%&-6.5955&-6.6306\\
			Monaghan County &-6.5670&-6.5733&-6.5732&-6.6326&-6.6341&-6.6311\\%&-6.5955&-6.6306\\
			Kildare County 	&-6.5638&-6.5733&-6.5732&-6.5689&-6.5674&-6.5645\\%&-6.5955&-6.6306\\
			Wicklow County 	&-6.5397&-6.5733&-6.5732&-6.5689&-6.5674&-6.5645\\%&-6.5955&-6.5627\\
			Wexford County 	&-6.5217&-6.5733&-6.5732&-6.5689&-6.5674&-6.5645\\%&-6.4897&-6.5627\\
			South Tipperary &-6.5062&-6.5000&-6.5023&-6.5006&-6.5674&-6.5645\\%&-6.4897&-6.5627\\
			Cavan County 	&-6.4809&-6.5000&-6.5023&-6.5006&-6.5011&-6.4980\\%&-6.4897&-6.5104\\
			Clare County 	&-6.4764&-6.5000&-6.5023&-6.5006&-6.5011&-6.4980\\%&-6.4897&-6.5104\\
			Cork County 	&-6.4738&-6.5000&-6.5023&-6.5006&-6.5011&-6.4980\\%&-6.4897&-6.5104\\
			Louth County 	&-6.4720&-6.5000&-6.5023&-6.5006&-6.5011&-6.4980\\%&-6.4897&-6.5104\\
			South Dublin 	&-6.4708&-6.5000&-6.5023&-6.5006&-6.5011&-6.4980\\%&-6.4897&-6.5104\\
			Dun Laoghaire-Rathdown&-6.4489&-6.4648&-6.4670&-6.4653&-6.4668&-6.4609\\%&-6.4501&-6.4732\\
			Limerick County &-6.4473&-6.4648&-6.4670&-6.4653&-6.4668&-6.4609\\%&-6.4501&-6.4732\\
			Cork City 		&-6.4385&-6.4648&-6.4670&-6.4653&-6.4668&-6.4609\\%&-6.4501&-6.4732\\
			Fingal 			&-6.4379&-6.4648&-6.4670&-6.4653&-6.4668&-6.4609\\%&-6.4501&-6.4732\\
			North Tipperary &-6.4323&-6.4648&-6.4670&-6.4653&-6.4668&-6.4609\\%&-6.4501&-6.4732\\
			Limerick City 	&-6.4306&-6.4648&-6.4670&-6.4653&-6.4668&-6.4609\\%&-6.4501&-6.4732\\
			Kilkenny County &-6.4299&-6.4648&-6.4670&-6.4653&-6.4668&-6.4609\\%&-6.4501&-6.4732\\
			Laoighis County &-6.3923&-6.3766&-6.3788&-6.3772&-6.3787&-6.4609\\%&-6.3865&-6.4732\\
			Carlow County 	&-6.3865&-6.3766&-6.3788&-6.3772&-6.3787&-6.3735\\%&-6.3865&-6.3859\\
			Longford County &-6.3813&-6.3766&-6.3788&-6.3772&-6.3787&-6.3735\\%&-6.3865&-6.3859\\
			Westmeath County&-6.3808&-6.3766&-6.3788&-6.3772&-6.3787&-6.3735\\%&-6.3865&-6.3859\\
			Dublin City 	&-6.3694&-6.3766&-6.3788&-6.3772&-6.3787&-6.3735\\%&-6.3865&-6.3859\\
			Galway City 	&-6.3421&-6.3766&-6.3788&-6.3772&-6.3787&-6.3735\\%&-6.3431&-6.3859\\
			Galway County 	&-6.3415&-6.3766&-6.3788&-6.3772&-6.3787&-6.3735\\%&-6.3431&-6.3859\\
			Kerry County 	&-6.3323&-6.3766&-6.3788&-6.3772&-6.3787&-6.3735\\%&-6.3431&-6.3859\\
			Meath County 	&-6.3282&-6.3766&-6.3788&-6.3772&-6.3787&-6.3735\\%&-6.3431&-6.3859\\
			Roscommon County&-6.3031&-6.3766&-6.3788&-6.3772&-6.3787&-6.3735\\%&-6.3431&-6.3859\\
			Sligo County 	&-6.2503&-6.2105&-6.2128&-6.2113&-6.2128&-6.2098\\%&-6.1956&-6.2221\\
			Leitrim County 	&-6.2282&-6.2105&-6.2128&-6.2113&-6.2128&-6.2098\\%&-6.1956&-6.2221\\
			Mayo County 	&-6.1615&-6.2105&-6.2128&-6.2113&-6.2128&-6.2098\\%&-6.1956&-6.2221\\
			\hline\hline
	\end{tabular} }
\end{table}

\begin{table}[htb]
	\centering
	\caption{The severity model coefficients for the standard GLM and the GLMs using factor collapsing. Categorical levels are of increasing order based on the standard GLM coefficients for ease of illustration. Among all selected models, the first 66 models and the first 130 models account for 80\% and 100\% model posterior probability respectively. Only the five best models are selected for ease of illustration. Note that for the five models shown only adjacent categorical levels are clustered in most cases. But as model weight becomes smaller, this will occur less often.}
	\label{table:county_coef_severity}
	\resizebox{\linewidth}{!}{
		\begin{tabular}{lrrrrrrrrrrrrrrrrrrrrr}
			\hline\hline
			\ &Standard GLM&Model 1&Model 2&Model 3&Model 4&Model 5\\%&Model 6&Model 7\\
			\hline
			BIC & \ &120,979.5446&120,979.5446&120,979.5446&120,979.5446&120,979.6222\\%&120979.6330&120979.7712\\
			Model weight among all models& \ &0.0159&0.0159&0.0159&0.0159&0.0159\\%&0.0159&0.0159\\
			\ \\
			Carlow County&8.4311&8.4111&8.4111&8.4111&8.4111&8.4182\\%&8.4043&8.3989\\
			South Dublin&8.4379&8.4111&8.4111&8.4111&8.4111&8.4182\\%&8.4043&8.3989\\
			Kildare County&8.4689&8.4111&8.4111&8.4111&8.4111&8.4182\\%&8.4043&8.3989\\
			Westmeath County&8.4727&8.4111&8.4111&8.4111&8.4111&8.4182\\%&8.4043&8.3989\\
			Limerick County&8.4900&8.4111&8.4111&8.4111&8.4111&8.4182\\%&8.4043&8.3989\\
			Roscommon County&8.4900&8.4111&8.4111&8.4111&8.4111&8.4182\\%&8.4043&8.3989\\
			Kilkenny County&8.5079&8.4760&8.4760&8.4760&8.4760&8.4824\\%&8.4710&8.3989\\
			Dublin City&8.5101&8.4760&8.4760&8.4760&8.4760&8.4824\\%&8.4710&8.4628\\
			Kerry County&8.5315&8.4760&8.4760&8.4760&8.4760&8.4824\\%&8.4710&8.4628\\
			Unknown&8.5321&8.4760&8.4111&8.5038&8.6392&8.4824\\%&8.6324&8.6192\\
			Leitrim County&8.4745&8.4760&8.4760&8.4760&8.4760&8.4824\\%&8.4710&8.4628\\
			Laois County&8.5367&8.4760&8.4760&8.4760&8.4760&8.4824\\%&8.4710&8.4628\\
			Cork City&8.5392&8.4760&8.4760&8.4760&8.4760&8.4824\\%&8.4710&8.4628\\
			North Tipperary&8.5436&8.4760&8.4760&8.4760&8.4760&8.4824\\%&8.4710&8.4628\\
			Offaly County&8.5441&8.4760&8.4760&8.4760&8.4760&8.4824\\%&8.4710&8.4628\\
			Wexford County&8.5456&8.4760&8.4760&8.4760&8.4760&8.4824\\%&8.4710&8.4628\\
			Dun Laoghaire-Rathdown&8.5611&8.5436&8.5436&8.5436&8.5436&8.5519\\%&8.5373&8.5247\\
			Waterford City&8.5776&8.5436&8.5436&8.5436&8.5436&8.5519\\%&8.5373&8.5247\\
			Meath County&8.5778&8.5436&8.5436&8.5436&8.5436&8.5519\\%&8.5373&8.5247\\
			Sligo County&8.5781&8.5436&8.5436&8.5436&8.5436&8.5519\\%&8.4710&8.4628\\
			Cork County&8.5840&8.5436&8.5436&8.5436&8.5436&8.5519\\%&8.5373&8.5247\\
			Fingal&8.5941&8.5436&8.5436&8.5436&8.5436&8.5519\\%&8.5373&8.5247\\
			Wicklow County&8.6035&8.5436&8.5436&8.5436&8.5436&8.5519\\%&8.5373&8.5247\\
			Galway City&8.6090&8.5436&8.5436&8.5436&8.5436&8.5519\\%&8.5373&8.5247\\
			Galway County&8.6185&8.5436&8.5436&8.5436&8.5436&8.5519\\%&8.5373&8.5247\\
			Mayo County&8.6207&8.5436&8.5436&8.5436&8.5436&8.5519\\%&8.5373&8.5247\\
			Monaghan County&8.6256&8.5436&8.5436&8.5436&8.5436&8.5519\\%&8.5373&8.5247\\
			Louth County&8.6321&8.5436&8.5436&8.5436&8.5436&8.5519\\%&8.5373&8.5247\\
			Donegal County&8.6371&8.5436&8.5436&8.5436&8.5436&8.5519\\%&8.5373&8.5247\\
			Cavan County&8.6434&8.5436&8.5436&8.5436&8.5436&8.5519\\%&8.5373&8.5247\\
			Limerick City&8.6487&8.6392&8.6392&8.6392&8.6392&8.5519\\%&8.6324&8.6192\\
			Clare County&8.6718&8.6392&8.6392&8.6392&8.6392&8.6518\\%&8.6324&8.6192\\
			Longford County&8.6749&8.6392&8.6392&8.6392&8.6392&8.6518\\%&8.6324&8.6192\\
			South Tipperary&8.7040&8.6392&8.6392&8.6392&8.6392&8.6518\\%&8.6324&8.6192\\
			Waterford County&8.7500&8.6392&8.6392&8.6392&8.6392&8.6518\\%&8.6324&8.6192\\
			\hline\hline
	\end{tabular} }
\end{table}

\section{Summary and discussion}
\label{sec:conclusion}

This article introduces the framework of factor collapsing with Bayesian model averaging (FC-BMA) for model selection and averaging within an insurance claim modelling setting, to overcome the problems of factor level selection when there are many categorical variables each with many levels and of model uncertainty. By applying the FC-BMA method to the Irish motor claims data set, it reveals geographical patterns of Irish counties that share similar risks. Interestingly, even though the models are much more parsimonious, with fewer predictors and fewer categories within the included factors, they still preform well on a stand alone basis or using model averaging, although the improvement realised by using BMA varies. 

Given variables to be collapsed or selected, FC-BMA essentially searches over the model space, either via an exhaustive search or a stochastic search, to find an optimal model region within which all models could be selected and averaged, so that model prediction will be improved. At the same time, because any factor with many categories is collapsed into one with a smaller number of levels, the reduction of the number of parameters also leads to an improvement in model parsimony and interpretability.
The stochastic optimisation process remains an active research field and other stochastic optimisation methods such as adaptive simulated annealing (\cite{Ingber2012}) and Bayesian optimization (\cite{Mockus2012}) might also provide satisfactory results. Further adaptations of other optimisation algorithms for factor collapsing could be considered.

The FC-BMA method introduced is mainly based within the GLM framework, in which interaction terms are often included. Since interaction terms can also be viewed as categorical variables, interaction collapsing will add another layer of uncertainty to the method. Although interaction terms could be collapsed using FC-BMA, caution should be taken to ensure the consistency and interpretability between collapsed main effects and collapsed interactions. When doing so, one downside is that given the large size of the data, having many interaction terms (i.e. many model parameters) in the baseline model before FC may adversely affect the computation time, or model fitting may not converge. It is suggested that FC is used only with main effects, after the optimal collapsing has been found, with interactions added based on this optimal collapsing to ensure the levels involving main effects and interactions are consistent. This also substantially reduces the number of parameters in the model before interactions are added and hence model interpretability is improved. 

Alternative applications of FC-BMA could be extended to other types of models such as general linear regression or, as is common practice in insurance pricing, to continuous variables that are banded to capture the non-linear relationship between rating factors and dependent variables. The FC method could also be used to find the optimal banding when controlling consecutive collapsings. For example treating policyholder age as an ordered categorical variable where each integer is a category, adjacent levels can be combined to create a more optimal banding even if it results in bands of unequal width. 

Factor collapsing is non-parametric in nature, as it uses brute force evaluation to check all possible subsets. 
%Therefore, we do not know any statistical properties of the final collapsing. 
So far we have been using a noninformative prior probability for each partition, which does not take other information in the data such as number of observations for each category into consideration.
Alternative means of setting the prior is a prime focus in the BMA literature. 
Hence further research in this area may lead to some improvement.
It is acknowledged that FC-BMA works best when the number of levels is less than 50. When there are more than 50 categories, it becomes a more typical multi-level factor problem where methods of keeping all categories unchanged such as mixed effects models should be considered.

%Here Dirichlet process or the Chinese restaurant problem is considered.

%in cases where number of levels is less than 50, even though a stochastic search might work to find a solution, but there may not be enough data for each level to have a good fit of GLM. Therefore, direct factor collapsing method might not be a very reasonable choice, unless some pre-specified conditions are made, such as certain groups are pre-specified to be together etc. 

%\textbf{Space partitioning}: this is a class of algorithm which partition a space and find the best partition
%\textbf{Partition refinement:} this is an algorithm that start from a single set, then refine this set step by step and find the best partition. We should use this as another idea for factor collapsing: instead of list all different ways of partition, we start from one and split this partition. This idea is similar to what I was thinking about Chinese restaurant process. 

\appendix

\section{The genetic algorithm}
\label{app:GA}

This appendix discusses implementing the genetic algorithm (GA), which is modified for use with the factor collapsing method. 
It uses three rules at each iteration to reproduce the next generation from the current population, namely the selection rule, the crossover rule and the mutation rule. 
\subsection{Selection rule} This rule decides how to select parents (i.e. collapsing combinations in factor collapsing) that contribute to producing the next generation based on their fitness (such as BIC). The fittest are more likely to be selected and paired to reproduce. 
There are various methods for selecting the fittest individuals, such as roulette wheel selection, Boltzman selection, tournament selection, rank selection and steady state selection (\cite{Mitchell1996}). Since the optimisation problem here is minimisation instead of maximisation, caution should be exercised for selection.  
%For example, in roulette wheel selection, the chance to be selected is proportional to the fitness of each individual, therefore the higher the fitness, the more chances they have.
%Rank selection selects individuals based on their ranking of fitness.
%Tournament selection, while slower and more complicated, can create more diverse population. It repeatedly samples the population and selects the fittest from each small sample. 
%Suppose you want to pick 20 individuals from 100. Randomly choose (with uniform probability) a small number of individuals (typically fewer than 10) from the 100 (with replacement). Keep the fittest one. Do this again and again until you have got your 20.
Elitism is often used within selection rules, which prevents the loss of the fittest already found to date by copying one or several of the best solutions directly to the new population. It can rapidly improve performance of the GA.

\subsection{Crossover rules} This rule decides how to combine two parents to reproduce offspring and how often crossover is performed.
It is performed by selecting one (or more) random parts within the length of the collapsing combinations and swapping selected parts, in the hope that offspring will have the good components of parents and better fitness.
There are many ways of encoding the algorithm to suit problems at hand such as binary encoding, permutation encoding and values encoding (\cite{Mitchell1996} and \cite{Whitley1994}).
%(see \url{http://www.obitko.com/tutorials/genetic-algorithms/encoding.php}).   
%Based on classical crossover implementation, some necessary changes are made to accommodate the FC method, which is show next.
In the FC case the graycode for set partition is used. Figure~\ref{fig:GA_crossover} presents examples of both individual collapsing and multiple collapsing.

\begin{figure}[H]
	\begin{minipage}{\linewidth}
		\resizebox{0.67\linewidth}{!}{
		\noindent$\mathcal{A} \left\{ \begin{array}{l}
		\mbox{Parent 1:} \dotuline{122324536} \hspace{0.3cm} \Rightarrow  \hspace{0.3cm} \mbox{Offspring 1:} \dotuline{12232}\underline{3425}  \\
		\mbox{Parent 2:} \underline{111213425} \hspace{0.3cm} \Rightarrow  \hspace{0.3cm} \mbox{Offspring 2:} \underline{11121}\dotuline{4536} \hspace{0.3cm} \Rightarrow  \hspace{0.3cm}  \underline{11121}\dotuline{3456}  \\
		\end{array} \right.
		$} \newline
		\resizebox{\linewidth}{!}{
		$
		\mathcal{B} \left\{ \begin{array}{l}
		\mbox{Parent 1:} \ \mathsf{\dotuline{12322} \| \dotuline{1223435} \| \dotuline{1234456} \| \dotuline{122324536} } \hspace{0.3cm} \Rightarrow \hspace{0.3cm} \mbox{Offspring 1:}  \ \mathsf{\dotuline{12322} \| \underline{1223345}\|\underline{1123245}\|\dotuline{122324536} } \\
		\mbox{Parent 2:} \ \mathsf{\underline{11231} \| \underline{1223345}  \| \underline{1123245} \| \underline{111213425}}  \hspace{0.3cm} \Rightarrow \hspace{0.3cm} \mbox{Offspring 2:} \  \mathsf{\underline{11231}\|\dotuline{1223435}\|\dotuline{1234456}\|\underline{111213425} } \\
		\end{array} \right.
		$ }
	\end{minipage} 
	\caption{$\mathcal{A}$ shows collapsing one factor of nine levels; the number and location of crossover points are chosen randomly. One crossover point at the 5th digit is chosen here as an example. Note the last step is to convert the graycode into the canonical format. $\mathcal{B}$ shows collapsing four factors; both parents are combinations of graycodes of the four factors. Crossover breaking points are between factors and can be chosen randomly as one, two or three in this case. Two crossover points have been chosen here as an example. }
	\label{fig:GA_crossover}
\end{figure}

\subsection{Mutation rule} This rule controls how and how often parts of the offspring are mutated
during reproduction by applying random changes to them. It represents random modification during evolution.
Mutation prevents falling into a local optimum when exploring the model space, but it should not occur very often, otherwise the GA will degenerate to a random search. 
%Generally, mutation rate should be very low. Best rates reported are about 0.5\%-1\%. 
Each combination consists of the graycode for each factor. 
When performing multiple collapsing, mutation changes a graycode by first choosing which graycode is to be mutated, then a random neighbour of a partition is generated as in Table~\ref{table:partition_neighbour_example}. 
Figure~\ref{fig:GA_mutation} shows an example of mutation with four rating factors.
When performing individual collapsing, mutation is simply choosing a neighbouring partition as in Table~\ref{table:partition_neighbour_example}.

\begin{figure}[H]
	\begin{minipage}{\linewidth}
		\begin{center}
			\begin{tabular}{lllll}
				Before mutation: $\mathsf{12322} \| \underline{\underline{\mathbf{1223435}}} \| \mathsf{1234456} \| \mathsf{122324536}$ \\
				\hspace{4.7cm}  $\Downarrow$  \\
				After mutation:\ \ \ $\mathsf{12322} \| \underline{\underline{\mathbf{1233435}}} \| \mathsf{1234456} \| \mathsf{122324536} $ \\		
			\end{tabular}
		\end{center}
	\end{minipage}
	\caption{Example of mutation with four collapsing factors: the 2nd graycode is mutated. The partition $1233435$ is one of the neighbours of $1223435$ by randomly changing one element (the third element).}
	\label{fig:GA_mutation}
\end{figure}

\section{Benchmarking the factor collapsing method versus alternatives}
\label{app:benchmark}

There are many methods in the literature that can be viewed as dealing with factor collapsing, as outlined in the introduction.
This section benchmarks the advantages of FC-BMA versus competing methods, using the Sweden TP insurance example in Section~\ref{sec:small_example_benchmark}.

\subsection{Pairwise multiple comparison}

Within the GLM framework, pairwise multiple comparison shows how similar any two levels are, based on all pairwise multiple comparisons under the general linear hypothesis as proposed in \textcite{Hothorn2008simultaneous}. 
It is an extention of multiple comparison in ANOVA models and involves considerations of multiple null hypotheses simultaneously and each hypothesis tests the equivalence of any pair of coefficients within a categorical factor.
If each of the null hypotheses is tested with type I error $\alpha$, then the overall type I error will be larger than $\alpha$ because of multiplicity and it becomes more likely that at least one null hypothesis will be mistakenly rejected. 
A common method to deal with this issue is Tukey's test (\cite{Bretz2011}). 
As it tests the equivalence of coefficients, it arises naturally for merging categorical levels and hence performs factor collapsing. However, it still utilises a single model approach without consideration of model uncertainty.

When implementing for all predictors, the results are as shown in Table~\ref{table:pairwise_multiple_comparison}. The collapsing combination does not look satisfactory, especially for severity where all predictors are collapsed to be excluded from the model. Comparison of predictions is shown in Table~\ref{table:example_prediction_comparison_frequency} and Table~\ref{table:example_prediction_comparison_severity}.

\begin{table}[H]
	\centering
	\caption{Results of all-pairwise multiple comparison for the frequency and severity models.}
	\label{table:pairwise_multiple_comparison}
	\begin{tabular}{llll}
		\hline
		\           & Frequency      & Severity \\
		\hline
		Kilometers  & (1)(2)(3)(4)(5)& (12345) \\
		Zone        & (12356)(47) & (1234567) \\
		Bonus       & (1)(2)(3)(4)(5)(6)(7) & (1234567) \\
		Make 		& (125789)(3)(4)(6)& (123456789) \\
		\hline
	\end{tabular}
\end{table}

\subsection{CART}

Classification and regression tree (CART) is a common method that creates a tree-like classification or regression model by recursively partitioning data (splitting predictors) into dichotomous segmentations and exhaustively searching all possible groupings (\cite{Friedman2001}).
In particular, when predictors are categorical at each node their categorical levels are partitioned based on homogeniety and the end nodes represent the final groupings, hence providing factor collapsing.
It is useful for exploring relationships in the absense of a good prior model and it also handles large data sets easily.
One characteristic of a regression tree is that it allows for interactions among predictors without specifying them, which can be a disadvantage. Another characteristic is that it may overfit the model. Similar to pairwise multiple comparison, it considers only one best model. 

\begin{landscape}
	\begin{figure}[h]
		\centering
		\includegraphics[width=18cm, height=12cm]{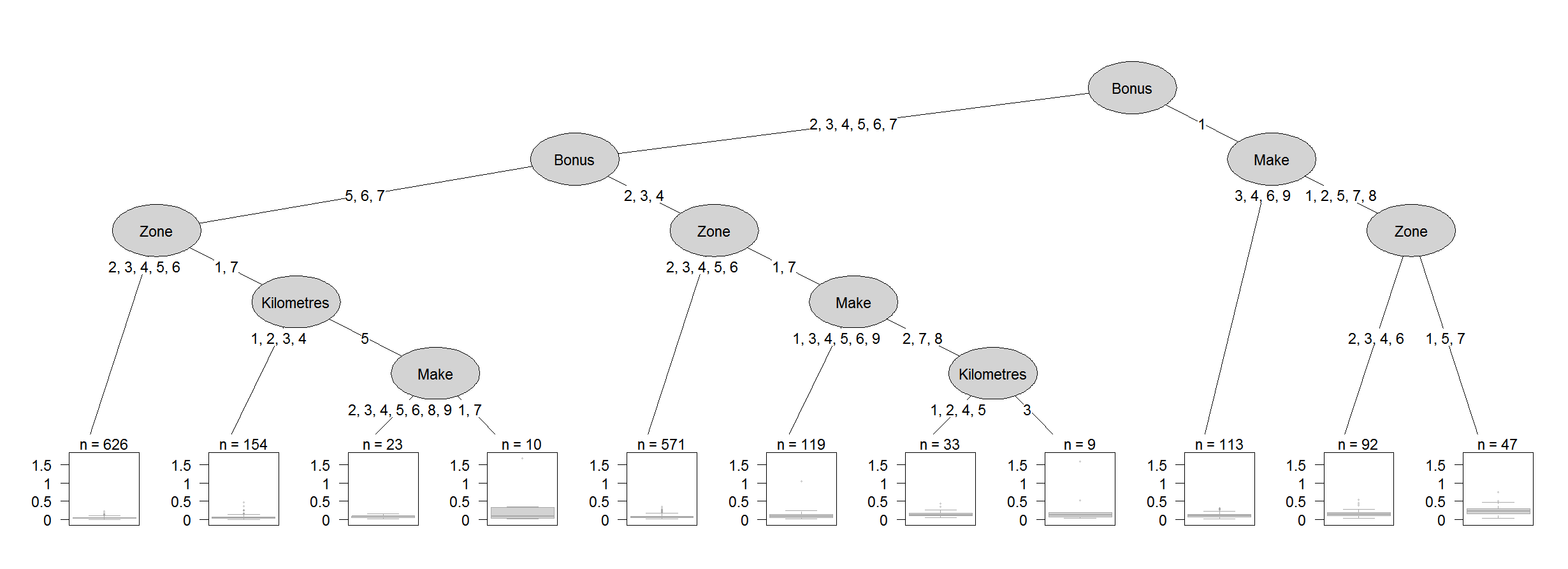}
		\caption{Plot of regression tree model for frequency for the Sweden TP motor insurance example.}
		\label{fig:tree_frequency}
	\end{figure}	
\end{landscape}

\begin{landscape}
	\begin{figure}[h]
		\centering
		\includegraphics[width=15cm, height=10cm]{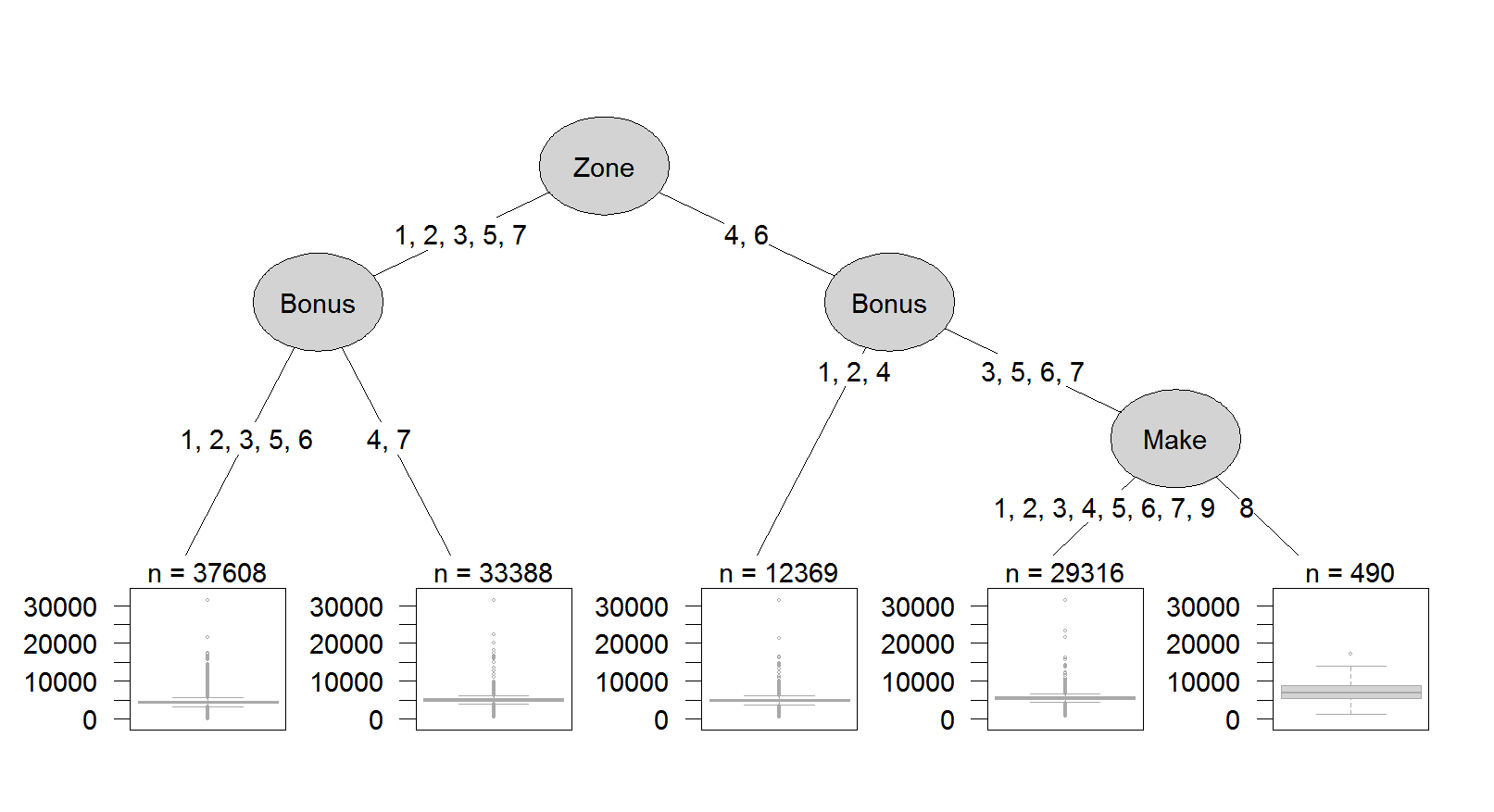}
		\caption{Plot of regression tree model for severity for the Sweden TP motor insurance example. }
		\label{fig:tree_severity}
	\end{figure}	
\end{landscape}

Regression tree models are implemented and the collapsing results are presented in Figure~\ref{fig:tree_frequency}, Figure~\ref{fig:tree_severity} and Table~\ref{table:tree_partition}.
The frequency tree model shows that there are interactions among all 4 predictors, which is consistent with the standard GLM model. But the final partitions are very different from the best partitions of FC in Table~\ref{table:4var_factor_collapsing_severity}. 
The severity tree model shows that Kilometers is not a significant predictor, there is interaction between Zone and Bonus, and again the final partitions are very different from those of FC. Note that the regression tree model does not provide model selection criteria such as BIC. Prediction accuracy is compared in Table~\ref{table:example_prediction_comparison_frequency} and Table~\ref{table:example_prediction_comparison_severity}.

\begin{table}[H]
	\centering	
	\caption{Partition of each rating factor based on regression tree model at each node and combined final partition }
	\begin{tabular}{l|llr}
		\hline
		\ & Predictor  &  Node partition & Final partition  \\ 
		\hline
		\multicolumn{1}{c|}{\multirow{9}{*}{Frequency} } & Kilometers &  (1245)(3) &  (124)(3)(5)  \\
		\multicolumn{1}{c|}{} & \          &  (1234)(5) &  \   \\
		\cline{2-4}
		\multicolumn{1}{c|}{} & Zone       &  (2346)(157) &  (17)(5)(2346)  \\  
		\multicolumn{1}{c|}{} & \          &  (23456)(17) &  \  \\
		\multicolumn{1}{c|}{} & \          &  (23456)(17) &  \  \\
		\cline{2-4}
		\multicolumn{1}{c|}{} & Bonus      &  (234567)(1) & (1)(234)(567)  \\ 
		\multicolumn{1}{c|}{} & \          &  (234)(567)  & \  \\
		\cline{2-4}
		\multicolumn{1}{c|}{} & Make       &  (3469)(12578)  & (1)(28)(3469)(5)(7)  \\
		\multicolumn{1}{c|}{} & \          &  (134569)(278)  & \  \\
		\multicolumn{1}{c|}{} & \          &  (2345689)(17)  & \  \\
		\hline 
		\multicolumn{1}{c|}{\multirow{4}{*}{Severity \ }} & Zone  &  (12357)(46)   & (12357)(46) \\
		\cline{2-4}
		\multicolumn{1}{c|}{} & Bonus &  (12356)(47)   & (12)(356)(4)(7) \\
		\multicolumn{1}{c|}{} & \     &  (124)(3567)   &  \  \\
		\cline{2-4} 
		\multicolumn{1}{c|}{} & Make  &  (12345679)(8) & (12345679)(8) \\
		\hline
	\end{tabular}
	\label{table:tree_partition}
\end{table}

\subsection{\emph{BMA} R package}

In the open-source software R (\cite{Rsoftware}), the \emph{BMA} package (\cite{BMApackage}) is widely used for performing Bayesian model averaging for model uncertainty in variable selection problems involving GLMs.
When correctly initialized it handles factor variables by independently and randomly selecting dummy variables of the model design matrix within each factor (i.e. each categorical level). This is equivalent to factor collapsing: by not selecting one level it means combining this unselected level with the baseline level. If interaction terms between factor variables are included, it will create dummy variables in the design matrix to represent each interaction between levels before all dummy variables are randomly selected. 
This leads to two differences compared to FC proposed in this article. First, the choice of baseline level in GLM fitting is important. All other levels except the reference level can only be merged with the reference level instead of being merged with categories other than the reference level. This gives a much smaller number of combinations, even when rotating different baseline choices. Second, the method for dealing with interactions makes it difficult and inconsistent in interpreting the factor collapsing results, especially when any of the corresponding main effects are dropped while being maintained in the interaction terms. 

For the Swedish TP frequency model with all two-way interactions included, 
there are 5 ``best" models and their cumulative posterior probability is one.
The best factor collapsing result can be interpreted as in Table~\ref{table:BMA_Rpackage_result}. 
The prediction is much worse than that of the standard GLM model, or the FC-BMA method.
If no interaction terms are included, then this method will return the standard GLM result. 
For the severity model, ``13" best models were selected by this method with cumulative posterior probability one and the ``best" 5 among these have cumulative posterior probability 0.821. The best collapsing result in Table~\ref{table:BMA_Rpackage_result} is very different from that of FC-BMA method. 
In both the frequency and severity models, predictions are worse using the \emph{BMA} package as shown in Table~\ref{table:example_prediction_comparison_frequency} and~\ref{table:example_prediction_comparison_severity}.

\begin{table}[H]
	\centering
	\caption{Results of \emph{BMA} R package for the frequency and severity models. Five models are selected for the frequency model and 13 models are selected for the severity model. The cumulative posterior probability is 1 for both cases. }
	\label{table:BMA_Rpackage_result}
	\begin{tabular}{llll}
		\hline\hline
		\ & Frequency & Severity \\
		\hline
		Kilometers  & (1)(2)(3)(4)(5)& (1234)(5) \\
		Zone        & (1)(2)(3)(4)(5)(6)(7)& (1347)(2)(5)(6) \\
		Bonus       & (1)(2)(3)(4)(5)(6)(7)& (1234567) \\
		Make 		& (1235679)(4)(8)&  (1245679)(3)(8) \\
		\hline\hline
	\end{tabular}
\end{table}

\subsection{Regularisation based methods}
Least absolute shrinkage and selection operator (lasso) was introduced by \textcite{Tibshirani1996} to improve the prediction accuracy and interpretability of regression models, as well as to perform variable selection.
It performs subset selection by posing a constraint of the form $\sum_{j=1}^p |\beta_j| \leq t$, where $\beta_j$ represents variable coefficients and $t$ is a free parameter to decide how much regularisation needed. 
%This requires the sum of the absolute value of the regression coefficients to be less than a fixed value and therefore forces certain coefficients to be set to zero. Theoretically when a coefficient is zero, it can be interpreted as being combined with the reference level, hence factor collapsing. 
%However, when there are correlated variables lasso lack the group effect and generally only picks one variable from a group of correlated variables.   
Further generalisation of lasso methods have since been proposed, including fused lasso (\cite{Tibshirani2005}),
%further imposes one more constraint of $\sum_{j=2}^p |\beta_j - \beta_{j-1}| \leq t_1$ to ensure coefficients changes in a smoother manner, but only consecutive coefficients. 
clustered lasso (\cite{She2010}) 
%is another generalisation to lasso that identifies and groups covariates based on their coefficients, by penalising the differences between the coefficients so that nonzero ones make clusters together, with one regularisation of the form $\sum_{i<j}^{p} |\beta_i - \beta_j| \leq t_2$. 
and pairwise fused lasso (\cite{Petry2011}).
In particular, \textcite{Gertheiss2010} proposed lasso-type sparse models for categorical variables, which control both variable exclusion and inclusion and also categorical level selection (i.e. factor collapsing). However, as with most sparse modeling, selection criteria have to be used to determine the optimal penalty parameter and it is still based on the one-best-model approach. When setting the penalty parameter to different values, multiple models could be obtained, but how to define the group of best models still remains unclear. This contrasts with the merits of the FC-BMA method: since it is non-parametric in nature, there are no extra parameters to be determined and model uncertainty can be considered systematically.

We note that at the time of writing, there is no available R package for performing total clustering based on regularisation approach. Hence a standard lasso is implmented, the results are shown in Table~\ref{table:regularisation_result}. 
\begin{table}[H]
	\centering
	\caption{Results of regularisation based lasso for the frequency and severity models.}
	\label{table:regularisation_result}
	\begin{tabular}{llll}
		\hline
		\ & Frequency & Severity \\
		\hline
		Kilometers  & (1)(2)(3)(4)(5)& (1)(2)(3)(4)(5) \\
		Zone        & (1)(2)(3)(4)(5)(6)(7)& (1)(2)(3)(4)(5)(6)(7) \\
		Bonus       & (1)(2)(3)(4)(5)(6)(7)& (1)(2)(3)(4)(5)(6)(7) \\
		Make 		& (178)(2)(3)(4)(5)(6)(9) &  (18)(2)(3)(4)(5)(6)(7)(9) \\
		\hline
	\end{tabular}
\end{table}

%================================================

%Shrinkage is necessary to deal with this problem, either by using penalized maximum likelihood estimation (e.g., R rms package's ols and lrm functions for quadratic (ridge) L2 penalty) or using random effects. You can get predictions for individual levels easily using penalized maximum likelihood estimation, or by using BLUPS in the mixed effects modeling context.

%So either reduce the levels by "hand" (creating a new "other" level for all those levels with insufficient data) or what about using L2/L1 regularisation

%one solution from the book 'Element of Statistical Learning'. The solution is mentioned in classification tree sections. Specifically, the solution orders the levels of the categorical predictor by the number of occurrence of each level in one class, and then treats the predictor as an ordered predictor. Perhaps the ordering approach only works for classification tree.  this approach is explained on page 329 of the book element of statistical learning 

% factor analysis or a discrete latent modeling technique. But majors are mutually exclusive categories, so I'm hesitant to exploit their covariance for anything

%\url{http://stats.stackexchange.com/questions/146907/principled-way-of-collapsing-categorical-variables-with-many-categories}
%\url{http://stats.stackexchange.com/questions/67938/how-to-handle-categorical-predictors-with-too-many-levels}

\section*{Acknowledgements}
This work was supported by the Science Foundation Ireland funded Insight Research Centre (SFI/12/RC/2289).

%\bibliographystyle{imsart-nameyear}
%\bibliography{PaperDraftArXivBib}

\clearpage
\addcontentsline{toc}{chapter}{\bibname} % Add an entry for the Bibliography in the Table of Contents
\printbibliography

\end{document}